\documentclass{article}
\usepackage{graphicx}
\usepackage{authblk}
\usepackage[numbers,sort&compress]{natbib}
\usepackage{amsmath}
\usepackage{graphicx}
\usepackage{subcaption}
\usepackage{float}
\usepackage{appendix}
\usepackage{bm}
\usepackage{cancel}
\usepackage[colorlinks=true,linktocpage=true,
linkcolor=blue,citecolor=blue]{hyperref}
\usepackage[a4paper]{geometry}
\usepackage{amssymb}
\usepackage{hyperref}
\begin{document}
\large
\title{\bf{Study of transport properties of a hot and dense QCD matter using a novel approximation method}}
\author[1]{Anowar Shaikh\thanks{anowar.19dr0016@ap.iitism.ac.in}}
\author[2]{Shubhalaxmi Rath\thanks{shubhalaxmi@academicos.uta.cl (Corresponding author)}}
\author[3]{Sadhana Dash\thanks{sadhana@phy.iitb.ac.in}}
\author[1]{Binata Panda\thanks{binata@iitism.ac.in}}
\affil[1]{Department of Physics\\ Indian Institute of Technology (Indian School of Mines) Dhanbad, Jharkhand 826004, India}
\affil[2]{Instituto de Alta Investigaci\'{o}n, Universidad de Tarapac\'{a}, Casilla 7D, Arica, Chile}
\affil[3]{Department of Physics\\ Indian Institute of Technology Bombay, Mumbai 400076, India}
\date{}
\maketitle
\begin{abstract}
We have studied the charge and the heat transport properties of a hot and dense QCD matter by solving the relativistic Boltzmann transport equation using a novel approximation method. Following the recently 
developed novel relaxation time approximation (RTA) model, we have proposed a novel Bhatnagar-Gross-Krook (BGK) model with a modified collision integral to carry out the aforementioned study. We have also compared our findings with the results of the novel RTA, the standard RTA and the standard BGK models. Our observation shows that the novel collision integrals for both the RTA and BGK models decrease the charge and the heat transport phenomena in the medium, as evidenced by the reduced values of the transport coefficients, such as the electrical conductivity and the thermal conductivity, when compared to the standard RTA and standard BGK models. Furthermore, certain observables associated with the abovementioned transport coefficients, such as the thermal diffusion constant and the Lorenz number have been explored using the novel approaches of the aforesaid models. We have found an overall decreasing trend of the thermal diffusion constant with the temperature in the novel BGK model, similar to the novel RTA model, but the magnitude remains higher throughout the temperature range. However, the magnitude of the thermal diffusion constant 
in the proposed novel BGK model remains conspicuously lower than its value in the standard BGK model. The magnitude of the Lorenz number in the novel BGK model remains significantly higher than that in the standard BGK model, but it is lower than that in the novel RTA model. We have also observed that the Lorenz number in all cases has an increasing trend at low temperatures, showing a violation of the Wiedemann-Franz law, whereas at high temperatures, it becomes saturated. The Lorenz number remaining above unity indicates that the thermal conductivity prevails over the electrical conductivity in the aforesaid models. 

\end{abstract}

\newpage

\section{Introduction}
A long time goal of the physics of ultrarelativistic heavy ion collisions is the observation of new states of matter predicted by the quantum chromodynamics (QCD). The experimental evidences of the relativistic heavy ion collider (RHIC) \cite{Skokov:2009qp} at BNL and the large hadron collider (LHC) \cite{Bzdak:2011yy} at CERN indicate a new state of the strongly interacting matter known as quark-gluon plasma (QGP). Following the discovery of QGP, various properties of the strongly coupled QCD matter became a topic of high interest. Different research groups have studied different features of QGP in recent years, such as the thermodynamic properties \cite{Moreau:2021clr, He:2022ywp}, the relativistic hydrodynamics \cite{Mitra:2020gdk, Mitra:2021owk, Mitra:2021ubx}, the heavy quark diffusion in QCD 
matter \cite{Scardina:2017ipo, Das:2010tj, Khowal:2021zoo}, the memory effect on heavy quark dynamics \cite{vinales2007anomalous, Das:2016llg}, the dissipative spin dynamics in relativistic matter \cite{Bhadury:2020cop, Bhadury:2020puc} etc. Nowadays, there is much interest in the phase diagram region at finite temperature and baryon chemical potential \cite{Aoki:2009sc, Aoki:2006we}. Many observations have also been made recently to investigate the impact of a strong magnetic field on the characteristics of hot QCD matter, such as the chiral magnetic effect \cite{Fukushima:2008xe, Kharzeev:2007jp}, the chiral vortical effect \cite{Kharzeev:2010gr}, the axial Hall current \cite{Pu:2014fva}, the refractive indices and decay constant \cite{Fayazbakhsh:2013cha,Fayazbakhsh:2012vr}, the photon and dilepton productions from QGP \cite{vanHees:2011vb,Shen:2013vja,Tuchin:2013bda}, the thermodynamic and magnetic properties \cite{Rath:2017fdv,Bandyopadhyay:2017cle,Rath:2018tdu,Karmakar:2019tdp,Rath:2020idp,Khan:2020rdw}, the axial magnetic effect \cite{Braguta:2014gea,Chernodub:2013kya}, the thermoelectric responses of hot QCD medium in a time varying magnetic field \cite{K:2021sct,K:2022pzc}, the magnetic and inverse magnetic catalysis \cite{Gusynin:1994re,Lee:1997zj}, the nonlinear electromagnetic current \cite{Kharzeev:2013ffa,Satow:2014lia} and the magnetohydrodynamics \cite{Roy:2015kma,Godunov:2015nea}. 

It is crucial to investigate different transport coefficients since they shed light on the genesis of the hot QCD medium. Among these transport coefficients, the thermal conductivity ($\kappa$) and the electrical conductivity ($\sigma_{el}$) characterize the movement of heat and charges in the medium, respectively. The hydrodynamic development of the strongly interacting matter at finite baryon density also depends on these two conductivities \cite{Kapusta:2012zb, Denicol:2012vq}. Therefore, significant effects on the observables at heavy ion collisions might result from changes in the charge and the heat transport coefficients.
There are several methods for determining these conductivities, such as the relativistic Boltzmann transport equation \cite{Muronga:2006zx, Puglisi:2014sha}, the Chapman-Enskog approximation \cite{Mitra:2016zdw,Mitra:2017sjo}, the lattice simulation \cite{Gupta:2003zh,Amato:PRL111'2013,Aarts:2014nba,Ding:LATTICE2014'2015,Ding:2016hua,Aarts:EPJA57'2021}, the correlator technique using Green-Kubo formula \cite{Nam:2012sg,Greif:2014oia,Feng:2017tsh} etc. Previous research has examined the impact of magnetic field on the electrical conductivity using a variety of methods, including the diluted instanton-liquid model \cite{Shuryak:ZPC38'1988,Shuryak:ARNPS47'1997,Nam:2012sg,Shuryak:PRD034023'2023,Shuryak:PRD034024'2023,Shuryak:PRD034025'2023}, the quenched SU(2) lattice gauge theory \cite{Buividovich:2010tn}, the effective fugacity technique \cite{Kurian:2017yxj}, the diagrammatic method utilizing the real-time formalism \cite{Hattori:2016cnt} etc. Studies on the electrical and the thermal conductivities of hot QCD matter in the presence of magnetic field within the kinetic theory approach have been carried out in references \cite{Kurian:2018qwb,Rath:2019vvi,Rath:2021ryd,Shaikh:2022sky}. 

The application of theoretical methodologies intended for systems of this type has been made viable 
by the strongly coupled fluid behavior of QGP. A simplified version of the kinetic equation with a 
simplified form of the collision term is frequently used because of various technical limitations. It is referred as the relaxation time approximation (RTA) \cite{Anderson:1974nyl}. In this model, the mean-free approach reduced the complexity of the collision term. The development in the nonrelativistic case has already been achieved by Bhatnagar-Gross-Krook (BGK) model \cite{Bhatnagar:1954zz} and Welander model \cite{welander1954temperature}. This approximation is utilized in many domains of physics and is recently 
used to investigate the hydrodynamization of the matter generated by the ultrarelativistic collisions of heavy ions. The relaxation time approximation has become essential for exploring the domain and microscopic basis of the relevance of hydrodynamics with relativistic effects. Although widely used, the approximation presented in ref. \cite{Anderson:1974nyl} has some fundamental challenges. It conflicts with some macroscopic and microscopic conservation laws. Recently, some works \cite{Rocha:2021zcw,Rocha:2021lze,Rocha:2022fqz} provided a novel relaxation time approximation that fixes such fundamental issues and maintains the core features of the linearized Boltzmann collision operator. In light of the momentum-dependent relaxation time, the novel relaxation time approximation is significant as it maintains the fundamental collision features associated with the microscopic conservation laws. They showed how this adjustment impacts the particle diffusion and the bulk viscosity. In a recent study \cite{Rath:2024vyq}, the transport coefficients related to charge and heat for a rotating QGP medium have been derived within the RTA model utilizing the aforementioned collision integral. 

In this study, we estimate the electrical conductivity ($\sigma_{el}$) and the thermal conductivity ($\kappa$), to investigate the charge and the thermal transport properties of a hot and dense QCD 
matter. These transport coefficients have been computed using a novel approximation framework that improves the standard RTA and standard BGK models significantly. The novel approaches of these models enhance the understanding of charge and heat transport phenomena in a strongly coupled thermal QCD medium while ensuring strict compliance with microscopic conservation laws. The findings of our study indicate that the modifications to the collision integral result in a substantial reduction of $\sigma_{\text{el}}$ and $\kappa$, implying that the transport of charge and heat in QGP is more constrained than earlier assessments based on the standard RTA and standard BGK models. This can lead to revised predictions for the QGP lifetime and photon/dilepton emission rates. We also explore the thermal diffusion constant and the Lorenz number as the applications of charge and heat transport properties in an analogous environment. The thermal diffusion constant shows the overall rate of heat transport in the medium, while the Lorenz number illustrates the connection between the thermal and electrical conductivities. We use the quasiparticle description of partons, where the rest masses are replaced by the masses generated in the medium. The quark-gluon plasma is viewed as a medium made up of thermally massive quasiparticles in this framework \cite{Braaten:1991gm,Peshier:2002ww}. 

The rest of the paper has been organized as follows. Using the modified collision integral, the charge and heat transport coefficients from the relativistic Boltzmann transport equation are calculated in 
section 2. In section 3, we have discussed the results of the aforesaid transport coefficients using the quasiparticle description of partons. Section 4 describes various applications and their results obtained 
from these coefficients, such as the Lorenz number and the thermal diffusion constant. In section 5, the results are summarized. 

\textbf{Notations and conventions}: The covariant derivatives $\partial_\mu$ and $\partial^{(p)} _\mu$ denote $\frac{\partial}{\partial x^\mu}$ and $\frac{\partial}{\partial p^\mu}$ respectively. The fluid four-velocity in the rest frame ($u^\mu=(1,0,0,0)$) is normalized to unity.  In this work, the subscript $f$ stands for the flavor index, with $f$=u, d, s. The symbols $q_f$, $g_f$ and $ \delta f_{f}$ ($\delta \bar {f}_{f}$) stand for electric charge, degeneracy factor and the infinitesimal change in the distribution function for the quark (antiquark) of the $f^{th}$ flavor, respectively. Here, $g_g=2(N^2_c-1)$ represents the degeneracy factor for the gluonic degree of freedom, and $g_f(g_{\bar{f}})=2N_c$ denotes the degeneracy factor for the quark (antiquark) degree of freedom with flavor $f$, where $N_c=3$ is the number of colours. 

\section{Charge and heat transport properties of a hot and dense QCD medium}
For a single particle distribution function, the relativistic Boltzmann transport equation (RBTE) is written as
\begin{equation}\label{65} 	
\frac{\partial f_f}{\partial t}+ \frac{\bf{p}}{m}\cdot\nabla f_f + \bf{F}\cdot\frac{\partial \textit{f}_\textit{f}}{\partial \bf{p}}=\left(\frac{{\partial \textit{f}_\textit{f}}}{\partial \textit{t}}\right)_c 
,\end{equation}
where \textbf{F} is the force field acting on the particles in the medium and $m$ is the mass of the particle. The term on the right-hand side results from particle collisions within the medium. This term's zero value 
denotes the Vlasov equation, which describes a collisionless system. The intricate character of the collision term presents a number of challenges even for the simplest solution of the above eq. \eqref{65}. The term $\left(\frac{{\partial \textit{f}_\textit{f}}}{\partial \textit{t}}\right)_c$ represents the instantaneous change in the distribution function due to collisions between particles. In order to arrive at a comprehensive solution, we will first study a few mathematical models those allow us to address the collision term in a suitable manner. 

When using the relaxation time approximation (RTA), the collision term is substituted with a relaxation term of the following form, 
	\begin{equation}\label{xp1}
		\left( \frac{\partial f_\textit{f}}{\partial t}\right) _c=-\frac{1}{\tau_{fp}}\Bigl( f_\textit{f}(x,p,t)-f_{eq,f}(p)\Bigr),
	\end{equation}
where we can write $f_\textit{f}(x,p,t)=f_{eq,f}(p)+\delta f_f(x,p,t)$ and $\tau_{fp}$ is the relaxation time for collisions, {\em i.e.} the time required to bring the perturbed system back to its equilibrium state. We can assume that the quark distribution function is near the equilibrium with a slight divergence from equilibrium in order to permit linearization of the relativistic Boltzmann transport equation. Here, the term $\frac{1}{\tau_{fp} }$ acts as a damping frequency. The initial issue with this model is that, there is no immediate conservation of charge. To remove the crunch, the Bhatnagar-Gross-Krook (BGK) model presents 
a new type of collision kernel, denoted as
\begin{equation}\label{xr2}
\left( \frac{\partial f_\textit{f}}{\partial t}\right)_c=-\frac{1}{\tau_{fp}}\left( f_f(x,p,t)-\frac {n_f (x,t)}{n_{eq, f}}f_{eq,f}(p)\right), 	
\end{equation}
where $n_{eq, f}$ represents the equilibrium particle number density and $n_f(x,t)$ is known as the perturbed particle number density or the fluctuating particle number density. The perturbed particle number  density is calculated as $\int_p f_f(x,p,t)$, which gives zero value after integrating over momentum, thus indicating instantaneous conservation of the particle number during collisions in the system. The fundamental problems regarding the nonconservation of macroscopic and microscopic conservation laws are resolved, and the essential characteristics of the linearized Boltzmann collision operator are preserved via a novel relaxation time approximation that was recently presented in a research \cite{Rocha:2021zcw}. 

Here, we are considering the general relativistic covariant form of the  Boltzmann transport equation for an isotropic medium of quarks, antiquarks and gluons, which is given as
\begin{equation}\label{pp4}
	\begin{split}
		p^\mu \partial_\mu f_f+q_fF^{\mu \nu} p_\nu \partial^{(p)} _\mu f_f =\textbf{C}\bigl[\textit{f}_\textit{f}\hspace{1mm}\bigr],
	\end{split}  
\end{equation}
where $F^{\mu \nu}$ is the electromagnetic field strength tensor. In order to see the effects of electric field, we take only $\mu=i$ and $\nu=0$ and vice versa components of the electromagnetic 
field strength tensor, i.e. $F^{i0}=\textbf{E}$ and $F^{0i}=-\textbf{E}$ in our calculation, such that the RBTE \eqref{pp4} takes the following form, 
	  \begin{equation}\label{D}
q_{f}\textbf{E}\cdot\textbf{p}\frac{\partial f_{eq,f}}{\partial p^0}+q_f p_0\textbf{E}\cdot\frac{\partial f_{eq,f}}{\partial \textbf{p}}=\textbf{C}\bigl[\textit{f}_\textit{f}\hspace{1mm}\bigr].
	  \end{equation}
The infinitesimal change in quark distribution function is denoted as $\delta f_f$  and $f_{eq,f}$ 
indicates the equilibrium distribution function in the medium for $f^{th}$ flavor. 

\subsection{Electrical conductivity in the novel RTA and the novel BGK models}
The electric current density of the medium must be determined in order to see how the electric field 
affects the charge transport phenomena. An external electric field can cause an infinitesimal 
disturbance to a thermal QCD medium that contains quarks and antiquarks of distinct flavors, and gluons. 
Due to this electric field, there is an induced current density in the medium, which is given by
	\begin{equation}\label{A}
		J^{i}=\sum_{f}  g_{f}q_{f}\int \frac {d^3p}{(2\pi)^3} \frac{p^i}{\omega_f}\bigl[\delta f_f(x,p)+\delta \bar{f}_f(x,p)\bigr],
	\end{equation}
where $J^{i}$ represents the spatial part of the electric current density vector. Ohm's law states that the spatial component of the electric current density is directly proportional to the electric field with a proportionality factor, known as the electrical conductivity ($\sigma_{el}$). Thus, we have 
\begin{equation}\label{C}
J^i=\sigma_{el}E^i
.\end{equation}
Comparing eq. \eqref{A} and eq. \eqref{C}, one can obtain the expression of the electrical conductivity. 

By solving the relativistic Boltzmann transport equation, the infinitesimal changes in the quark and antiquark distribution functions can be obtained. The relativistic Boltzmann transport equation in the novel RTA model can be written as
	\begin{equation}\label{cc1}
	q_{f} \textbf{E}\cdot\textbf{p}\frac{\partial f_{eq,f}}{\partial p^0}+q_f p_0\textbf{E}\cdot\frac{\partial f_{eq,f}}{\partial \textbf{p}}=\textbf{C}\bigl[\textit{f}_\textit{f}\bigr],
	\end{equation}
where the collision integral is given \cite{Rocha:2021zcw} by
\begin{multline}\label{xy8}	
	\textbf{C}\bigl[\textit{f}_\textit{f}\hspace{1mm}\bigr]=-\frac{\omega_f}{\tau_{fp}}\biggr[\delta f_f-
	\frac{\left\langle({\omega_f}/{\tau_{fp}})\delta f_f\right\rangle_0}{\left\langle{\omega_f}/{\tau_{fp}}\right\rangle_0}+
	P^{(0)}_1\frac{\left\langle({\omega_f}/{\tau_{fp}})P^{(0)}_1\delta f_f\right\rangle_0}{\left\langle({\omega_f}/{\tau_{fp}})P^{(0)}_1P^{(0)}_1\right\rangle_0}\\ 
	+ p^{\left\langle\mu\right\rangle}\frac{\left\langle({\omega_f}/{\tau_{fp}})p_{\left\langle\mu\right\rangle}\delta f_f\right\rangle_0}{(1/3)\left\langle({\omega_f}/{\tau_{fp}})p_{\left\langle\mu\right\rangle}p^{\left\langle\mu\right\rangle}\right\rangle_0}\Biggl].	
\end{multline}
Here, 
\begin{equation}
	P^{(0)}_1=1-\frac{\left\langle{\omega_f}/{\tau_{fp}}\right\rangle_0}{\left\langle{\omega^2_f}/{\tau_{fp}}\right\rangle_0}\omega_f,
\end{equation}
\begin{equation}
p^{\left\langle\mu\right\rangle}=\Delta^{\mu\nu}p_{\nu}	,
\end{equation}
\begin{equation}
\Delta^{\mu\nu}=g^{\mu\nu}-u^{\mu}u^{\nu},	
\end{equation}
and we define the momentum integrals relative to the local equilibrium distribution function $f_{eq,f}$ as follows, 
\begin{eqnarray}
\nonumber\left\langle ...\right\rangle_0 &=& \int \frac{d^3p}{(2\pi)^3p_0} f_{eq,f}(...) \\ &=& \int dP f_{eq,f}(...)
,\end{eqnarray}
where $\tau_{fp}$ is the relaxation time and $\nu_f=\frac{1}{\tau_{fp}}$ denotes the collision frequency 
of the medium. The equilibrium distribution functions for $f^{th}$ flavor of quark and antiquark are 
given by
\begin{eqnarray}
&&f_{eq,f}=\frac{1}{e^{\beta(\omega_f-\mu_f)}+1}, \\ 
&&\bar{f}_{eq,f}=\frac{1}{e^{\beta(\omega_f+\mu_f)}+1}
,\end{eqnarray}
respectively, where $\omega_f=\sqrt{\textbf{p}^2+m^2_f}$, $\beta=\frac{1}{T}$, and $\mu_f$ represents the chemical potential of $f^{th}$ flavor of quark. Getting inspired by references \cite{Dusling:2009df, Dusling:2011fd, Kurkela:2017xis}, we have used the relaxation time of the following form, 
\begin{equation}\label{p77}
\tau_{fp}(\omega_f)=(\beta \omega_f)^{\gamma} \tau_{f}
.\end{equation}
Here, $\gamma$ is an arbitrary constant which controls the energy dependence of the relaxation time and has different values for different theories, for example, in QCD kinetic theories, $\gamma=\frac{1}{2}$ \cite{Dusling:2009df} and in scalar field theories $\gamma=1$ \cite{Calzetta:PRD37'1988}. Many previously published studies on QGP transport characteristics had postulated a power-law relation 
between the relaxation time and the energy, with $\gamma=0.5$ \cite{Dusling:2009df,Dusling:2011fd,Kurkela:2017xis}. By including the scale dependency of the relaxation time ($\tau_{fp}$), we enhance the novel RTA approximation with features known in QCD. The choice of $\gamma$ guarantees that the transport coefficients such as electrical conductivity, shear viscosity and thermal conductivity align with anticipated behaviors in quark-gluon plasma. In the analogous QCD scenario, high energy particles mainly lose energy through inelastic gluon radiation. The relaxation time scales as $\tau_{f p}\propto E^{1/2}$ due to the Landau-Pomeranchuk-Migdal suppression effect. The energy loss rate of a high energy parton follows $\frac{dE}{dt} \sim E^{1/2}$, which means that it increases with energy, but at a sublinear rate. An increased value of $\gamma$ (e.g., $\gamma=1$) would result in an excessively fast increase of the relaxation time with energy, hence causing an overestimation of the transport coefficients. A smaller value of $\gamma$ (e.g., $\gamma=0$) might make it insufficiently reliant on energy, which contradicts the anticipated quark scattering behavior in QGP. Thus, we set the value of $\gamma$ to $0.5$ in the present work. In the above equation, $\tau_f$ is momentum-independent, whose expression is given \cite{Hosoya:1983xm} by
\begin{equation}\label{1e}
\tau_{f}=\frac{1}{5.1T\alpha^2_s \log(1/\alpha_s)\left[1+0.12(2N_f+1)\right]}
,\end{equation}
where $\alpha_s$ is the QCD running coupling constant, which is a function of both temperature and chemical potential, and it has the following \cite{kapusta1989finite} form, 
\begin{eqnarray}\label{ww2}
\alpha_s=\frac{g^2}{4\pi}=\frac{12\pi}{\left(11N_c-2N_f\right)\ln\left({\Lambda^2}/{\Lambda_{\rm\overline{MS}}^2}\right)}
~.\end{eqnarray}
Here, $\Lambda_{\rm\overline{MS}}$ is the renormalization scale parameter in the modified minimum subtraction (MS) scheme, a widely used QCD renormalization approach in the determination of the QCD 
running coupling constant. In case of the high temperature QGP medium, one can use $\Lambda_{\rm\overline{MS}}=0.176$ GeV for one-loop running coupling constant as per lattice measurements \cite{Bazavov:PRD86'2012} and particle data group \cite{Beringer:PRD86'2012}. In the above equation, the renormalization scale $\Lambda=2\pi\sqrt{T^2+\mu_f^2/\pi^2}$ for electrically charged particles (quarks and antiquarks) and $\Lambda=2 \pi T$ for gluons. Using the expression of $\textbf{C}\bigl[\textit{f}_\textit{f}\bigr]$ in eq. \eqref{cc1}, we have 
\begin{multline}\label{D9}
	q_{f}\textbf{E}\cdot\textbf{p}\frac{\partial f_{eq,f}}{\partial p^0}+q_f p_0\textbf{E}\cdot\frac{\partial f_{eq,f}}{\partial \textbf{p}}=-\frac{\omega_f}{\tau_{fp}}\biggr[\delta f_f-
	\frac{\left\langle({\omega_f}/{\tau_{fp}})\delta f_f\right\rangle_0}{\left\langle{\omega_f}/{\tau_{fp}}\right\rangle_0} \\ +P^{(0)}_1\frac{\left\langle({\omega_f}/{\tau_{fp}})P^{(0)}_1\delta f_f\right\rangle_0}{\left\langle({\omega_f}/{\tau_{fp}})P^{(0)}_1P^{(0)}_1\right\rangle_0}+p^{\left\langle\mu\right\rangle}\frac{\left\langle({\omega_f}/{\tau_{fp}})p_{\left\langle\mu\right\rangle}\delta f_f\right\rangle_0}{(1/3)\left\langle({\omega_f}/{\tau_{fp}})p_{\left\langle\mu\right\rangle}p^{\left\langle\mu\right\rangle}\right\rangle_0}\Biggl]
.\end{multline}
Now, solving the above equation one can extract the expression of $\delta f_f$ (for a detailed calculation see appendix \ref{appendix A}) as
	\begin{equation}\label{kl01}
\delta f_f=\frac{2q_f\beta\tau_{fp} (\textbf{E}\cdot\textbf{p}) f_{eq,f}(1-f_{eq,f})}{\omega_fA},	
\end{equation}
where 
\begin{multline}
	A=\left[1-\frac{\omega_f\int p^2 (f_{eq,f}/ \tau_{fp}) dp}{\int p^2\omega_f (f_{eq,f}/\tau_{fp})dp}\right]\frac{\int p^2 (f_{eq,f}/\tau_{fp})\Bigr[1-\Bigr(\frac{\int p^2 (f_{eq,f}/\tau_{fp})dp}{\int p^2 \omega_f (f_{eq,f}/\tau_{fp})dp}\Bigl)\omega_f\Bigl] dp}{\int p^2 (f_{eq,f}/\tau_{fp})\Bigr[1-\Bigr(\frac{\int p^2 (f_{eq,f}/\tau_{fp})dp}{\int p^2 \omega_f (f_{eq,f}/\tau_{fp})dp}\Bigl)\omega_f\Bigl]^2 dp}\\ 
	+\frac{3p\int p^3 (f_{eq,f}/ \tau_{fp}) dp}{\int p^4 (f_{eq,f}/ \tau_{fp}) dp}.	
\end{multline}
Similarly for antiquarks, we have 
\begin{equation}\label{kl02}
\delta {\bar{f}}_f=\frac{2q_ {\bar{f}}\beta\tau_ {\bar{fp}} (\textbf{E}\cdot\textbf{p}) {\bar{f}}_{eq,f}(1- {\bar{f}}_{eq,f})}{\omega_f\bar{A}},
\end{equation}
where 
\begin{multline}
\bar{A}=	\left[1-\frac{\omega_f\int p^2 ({\bar{f}}_{eq,f}/ \tau_ {\bar{fp}}) dp}{\int p^2\omega_f ({\bar{f}}_{eq,f}/\tau_ {\bar{fp}})dp}\right]\frac{\int p^2 ({\bar{f}}_{eq,f}/\tau_ {\bar{fp}})\Bigr[1-\Bigr(\frac{\int p^2 ({\bar{f}}_{eq,f}/\tau_ {\bar{fp}})dp}{\int p^2 \omega_f ({\bar{f}}_{eq,f}/\tau_ {\bar{fp}})dp}\Bigl)\omega_f\Bigl] dp}{\int p^2 ({\bar{f}}_{eq,f}/\tau_ {\bar{fp}})\Bigr[1-\Bigr(\frac{\int p^2 ({\bar{f}}_{eq,f}/\tau_ {\bar{fp}})dp}{\int p^2 \omega_f ({\bar{f}}_{eq,f}/\tau_ {\bar{fp}})dp}\Bigl)\omega_f\Bigl]^2 dp}\\ 
+\frac{3p\int p^3 ({\bar{f}}_{eq,f}/ \tau_ {\bar{fp}}) dp}{\int p^4 ({\bar{f}}_{eq,f}/ \tau_ {\bar{fp}}) dp}
.\end{multline}
Finally, replacing the values of $\delta f_f$ and $\delta {\bar{f}}_f$ in eq. (\ref{A}) and comparing it with eq. (\ref{C}), we get the final expression of the electrical conductivity in the novel RTA model as
	\begin{equation}\label{oo1}	
	\sigma^{Novel-RTA}_{el}=\frac{\beta}{3\pi^2}\sum_{f} g_f q^2_f\int dp\frac{p^4}{\omega^2_f } \Biggr[\frac{\tau_{fp}  f_{eq,f}(1-f_{eq,f})}{A}+\frac{\tau_ {\bar{fp}}   {\bar{f}}_{eq,f}(1- {\bar{f}}_{eq,f})}{\bar{A}}\Biggl].
\end{equation}

In the standard BGK model, the relativistic Boltzmann transport equation is written as
\begin{equation}\label{rf}
	q_{f}\textbf{E}\cdot\textbf{p}\frac{\partial f_{eq,f}}{\partial p^0}+q_f p_0\textbf{E}\cdot\frac{\partial f_{eq,f}}{\partial \textbf{p}}=\textbf{C}\bigl[\textit{f}_\textit{f}\hspace{1mm}\bigr]=-\frac{p^\mu u_\mu}{\tau_{fp}}\left( f_f(x,p,t)-\frac {n_f(x,t)}{n_{eq, f}}f_{eq,f}(p)\right)
,\end{equation}
where the collision integral is given as
\begin{equation}
    \textbf{C}\bigl[\textit{f}_\textit{f}\hspace{1mm}\bigr]=-\frac{p^\mu u_\mu}{\tau_{fp}}\left( f_f(x,p,t)-\frac {n_f(x,t)}{n_{eq, f}}f_{eq,f}(p)\right)
.\end{equation}
In the above equation, $n_f$ and $n_{eq,f}$ represent the perturbed particle number density and the equilibrium particle number density, respectively for the $f^{th}$ flavor, which has a degeneracy 
of $g_f$. Both can be calculated from the equations as given below, 
\begin{equation}
	n_f=g_f\int dP \omega_f (f_{eq,f}+\delta f_f),
\end{equation}
\begin{equation}\label{ck}
	n_{eq,f}=g_f\int dP \omega_f  f_{eq,f}.
\end{equation}
The BGK collision kernel's instantaneous conservation of the particle number is its main 
benefit over the RTA collision kernel 
\cite{Bhatnagar:1954zz, Khan:2020rdw, Schenke:2006xu}, {\em i.e.}, in case of the BGK 
model, we have
\begin{equation}\label{rf6}
	\int dP \textbf{C}\bigl[\textit{f}_\textit{f}\hspace{1mm}\bigr]=0.
\end{equation}
Furthermore, the ref. \cite{Manuel:2004gk} had shown that, by implementing a straightforward adjustment, the BGK collision kernel effectively maintains the conservation of the color current in a covariant 
manner. The Boltzmann equation has been recently formulated using the BGK collision kernel to represent relativistic dissipative hydrodynamics \cite{Singha:2023eia}. 

In the present work, we proposed a collision integral in the following manner, 
\begin{multline}\label{rf1}
\textbf{C}\bigl[\textit{f}_\textit{f}\hspace{1mm}\bigr]^{NBGK}= -\frac{p^\mu u_\mu}{\tau_{fp}}\Bigg[\delta f_f-
	\frac{\left\langle({\omega_f}/{\tau_{fp}})\delta f_f\right\rangle_0}{\left\langle{\omega_f}/{\tau_{fp}}\right\rangle_0}+
	P^{(0)}_1\frac{\left\langle({\omega_f}/{\tau_{fp}})P^{(0)}_1\delta f_f\right\rangle_0}{\left\langle({\omega_f}/{\tau_{fp}})P^{(0)}_1P^{(0)}_1\right\rangle_0}
  \\
	+ p^{\left\langle\mu\right\rangle}\frac{\left\langle({\omega_f}/{\tau_{fp}})p_{\left\langle\mu\right\rangle}\delta f_f\right\rangle_0}{(1/3)\left\langle({\omega_f}/{\tau_{fp}})p_{\left\langle\mu\right\rangle}p^{\left\langle\mu\right\rangle}\right\rangle_0}\Bigg]
 + g_f n^{-1}_{eq,f} (p^\mu u_\mu) f_{eq,f}\int \frac{p^\mu u_\mu}{\tau_{fp}}\Bigg[ \delta f_f-
	\frac{\left\langle({\omega_f}/{\tau_{fp}})\delta f_f\right\rangle_0}{\left\langle{\omega_f}/{\tau_{fp}}\right\rangle_0}\\
 +
	P^{(0)}_1\frac{\left\langle({\omega_f}/{\tau_{fp}})P^{(0)}_1\delta f_f\right\rangle_0}{\left\langle({\omega_f}/{\tau_{fp}})P^{(0)}_1P^{(0)}_1\right\rangle_0}
	+ p^{\left\langle\mu\right\rangle}\frac{\left\langle({\omega_f}/{\tau_{fp}})p_{\left\langle\mu\right\rangle}\delta f_f\right\rangle_0}{(1/3)\left\langle({\omega_f}/{\tau_{fp}})p_{\left\langle\mu\right\rangle}p^{\left\langle\mu\right\rangle}\right\rangle_0}\Bigg] dP.
\end{multline}
To fulfill the basic conservation equations in microscopic interactions, the zeroth and first moments of the collision kernel must vanish (detailed calculations are given in appendix \ref{appendix B}). 

In the novel BGK model, the relativistic Boltzmann transport equation can be expressed as
\begin{multline}
	q_{f}\textbf{E}\cdot\textbf{p}\frac{\partial f_{eq,f}}{\partial p^0}+q_f p_0\textbf{E}\cdot\frac{\partial f_{eq,f}}{\partial \textbf{p}}= -\frac{p^\mu u_\mu}{\tau_{fp}}\Bigg[\delta f_f-
	\frac{\left\langle({\omega_f}/{\tau_{fp}})\delta f_f\right\rangle_0}{\left\langle{\omega_f}/{\tau_{fp}}\right\rangle_0}+
	P^{(0)}_1\frac{\left\langle({\omega_f}/{\tau_{fp}})P^{(0)}_1\delta f_f\right\rangle_0}{\left\langle({\omega_f}/{\tau_{fp}})P^{(0)}_1P^{(0)}_1\right\rangle_0}
  \\
	+ p^{\left\langle\mu\right\rangle}\frac{\left\langle({\omega_f}/{\tau_{fp}})p_{\left\langle\mu\right\rangle}\delta f_f\right\rangle_0}{(1/3)\left\langle({\omega_f}/{\tau_{fp}})p_{\left\langle\mu\right\rangle}p^{\left\langle\mu\right\rangle}\right\rangle_0}\Bigg]
 + g_f n^{-1}_{eq,f} (p^\mu u_\mu) f_{eq,f}\int \frac{p^\mu u_\mu}{\tau_{fp}}\Bigg[ \delta f_f-
	\frac{\left\langle({\omega_f}/{\tau_{fp}})\delta f_f\right\rangle_0}{\left\langle{\omega_f}/{\tau_{fp}}\right\rangle_0}\\
 +
	P^{(0)}_1\frac{\left\langle({\omega_f}/{\tau_{fp}})P^{(0)}_1\delta f_f\right\rangle_0}{\left\langle({\omega_f}/{\tau_{fp}})P^{(0)}_1P^{(0)}_1\right\rangle_0}
	+ p^{\left\langle\mu\right\rangle}\frac{\left\langle({\omega_f}/{\tau_{fp}})p_{\left\langle\mu\right\rangle}\delta f_f\right\rangle_0}{(1/3)\left\langle({\omega_f}/{\tau_{fp}})p_{\left\langle\mu\right\rangle}p^{\left\langle\mu\right\rangle}\right\rangle_0}\Bigg] dP.
	\end{multline}
Now, neglecting higher order terms $\left(\mathcal O(\delta f_f)^2\right)$, we get the solution of the above equation as
\begin{equation}\label{dr1}
\delta f_f=\delta f^{(0)}_f +g_f n^{-1}_{eq,f} f_{eq,f}\int_{ p\prime } \delta f^{(0)}_f \footnote{Here, we are using the momentum integration symbol, $\int_{p}=\int {d^3p}/{(2\pi)^3}$.}
,\end{equation}
where $\delta f^{(0)}_f$ is given by
\begin{equation}
	\delta f^{(0)}_f=\frac{2q_f\beta\tau_{fp} (\textbf{E}\cdot\textbf{p}) f_{eq,f}(1-f_{eq,f})}{\omega_fA}.
\end{equation}
The infinitesimal change in the quark distribution function is obtained (details are given in appendix \ref{appendix C}) as
\begin{equation}\label{xx2}
	\delta f_f=\Bigl[ 2q_f\beta  \tau_{fp} \frac{\textbf{E}\cdot\textbf{p}} {\omega_f A}f_{eq,f}(1-f_{eq,f}) \Bigr]
	+ g_f n^{-1}_{eq,f} f_{eq,f} \int_{p \prime} \Bigr[ 2q_f\beta  \tau_{fp}\frac{\textbf{E}\cdot\textbf{p}} {\omega_f A} f_{eq,f}(1-f_{eq,f})
	 \Bigl]. 
\end{equation}
Similarly, $\delta {\bar{f}}_f$ is obtained as
\begin{equation}
		\delta {\bar{f}}_f=\Bigl[ 2q_{\bar{f}}\beta  \tau_ {\bar{fp}} \frac{\textbf{E}\cdot\textbf{p}} {\omega_f \bar{A}}{\bar{f}}_{eq,f}(1-{\bar{f}}_{eq,f}) \Bigr]
	+ g_f n^{-1}_{eq,f} {\bar{f}}_{eq,f} \int_{p \prime} \Bigr[ 2q_{\bar{f}}\beta  \tau_ {\bar{fp}}\frac{\textbf{E}\cdot\textbf{p}} {\omega_f \bar{A}} {\bar{f}}_{eq,f}(1-{\bar{f}}_{eq,f})
	 \Bigl]. 
\end{equation}
Now, replacing the values of $\delta f_f$ and $\delta {\bar{f}}_f$ in eq. (\ref{A}) and comparing it with eq. (\ref{C}), we get the final expression of the electrical conductivity in the novel BGK model as
\begin{multline}\label{xx3}
	\sigma^{Novel-BGK}_{el}=\frac{\beta}{3\pi^2}\sum_{f} g_f q^2_f\int dp\frac{p^4}{\omega^2_f}\Biggl[\frac{\tau_{fp}f_{eq,f}(1-f_{eq,f})}{A}+\frac{\tau_ {\bar{fp}} {\bar{f}}_{eq,f}(1-{\bar{f}}_{eq,f})}{\bar{A}}
	 \Biggr]\\ +   
	\frac{2\beta}{\sqrt{3}}\sum_{f} g^2_f q^2_f n^{-1}_{eq,f}\Biggl[\int_{p} \frac{p}{\omega_f} f_{eq,f}  \int_{p \prime}\frac{p \prime}{\omega_f} \frac{\tau_{fp}f_{eq,f}(1-f_{eq,f})}{A}+\int_{p} \frac{p}{\omega_f} {\bar{f}}_{eq,f} \int_{p \prime}\frac{p \prime}{\omega_f}\frac{\tau_ {\bar{fp}} {\bar{f}}_{eq,f}(1-{\bar{f}}_{eq,f})}{\bar{A}}\Biggr]
.\end{multline}

\subsection{Thermal conductivity in the novel RTA and the novel BGK models}
Heat flow results from the deviation of the system from its equilibrium condition caused by the presence 
of a temperature gradient. Through a proportionality constant, {\em i.e.} the thermal conductivity, the 
heat flow vector is directly proportional to the temperature gradient. The calculation of the thermal conductivity is necessary in examining the heat transmission in a medium. In four-vector notation, the heat flow is written as
	\begin{equation}
		Q_{\mu}=\Delta_{\mu \alpha}T^{\alpha\beta}u_\beta-h\Delta_{\mu \alpha}N^\alpha.
	\end{equation}
In the above equation, the projection operator ($\Delta_{\mu \alpha}$) and the enthalpy per particle ($h$) are respectively defined as
	\begin{equation}
		\Delta_{\mu \alpha}=g_{\mu \alpha}-u_{\mu} u_{\alpha},
	\end{equation}
	\begin{equation}\label{pp6}
		h=(\varepsilon+P)/n,
	\end{equation}   	
where $n$, $\varepsilon$ and $P$ represent the particle number density, the energy density and the pressure, respectively, which are respectively given by the following equations, 
\begin{equation}
n=N^{\alpha}u_{\alpha}
,\end{equation}
\begin{equation}\label{E}
\varepsilon=u_{\alpha}T^{\alpha \beta}u_{\beta}
,\end{equation}
\begin{equation}\label{P}
P=-\frac{\Delta_{\alpha\beta} T^{\alpha\beta}}{3}
.\end{equation}
Here, the particle flow four-vector $N^{\alpha}$	and the energy-momentum tensor $T^{\alpha \beta}$ are respectively defined as
	\begin{equation}
		N^{\alpha}=\sum_{f} g_f \int \frac{d^3p}{(2\pi)^3}\frac{p^{\alpha}}{\omega_f}\bigl[f_f(x,p)+\bar{f}_f(x,p)\bigr],	
	\end{equation}
	\begin{equation}
		T^{\alpha \beta}=\sum_{f} g_f \int \frac{d^3p}{(2\pi)^3} \frac{p^{\alpha} p^{\beta}}{\omega_f}\bigl[f_f(x,p)+\bar{f}_f(x,p)\bigr]. 	
	\end{equation}
	The spatial part of the heat flow four-vector is given by
	\begin{equation}\label{K}
		Q^{i}=\sum_{f} g_f \int \frac{d^3p}{(2\pi)^3}\frac{p^i}{\omega_f}\bigl[(\omega_f -h_f)\delta f_f(x,p)+(\omega_f -\bar {h}_f)\delta \bar{f}_f(x,p)\bigr],
	\end{equation} 
where we use the fact that in the rest frame of the fluid, the heat flow four-vector and fluid four-vector are perpendicular to each other, {\em i.e.} $Q_{\mu}u^{\mu}=0$ and the enthalpy per particle for the $f^{th}$  flavor is given by
\begin{equation}
h_f=\frac{(\varepsilon_{eq,f}+P_{eq,f})}{n_{eq,f}}
,\end{equation}
where 
\begin{equation}\label{bb}
\varepsilon_{eq,f}=g_f\int dP~ \omega^2_f f_{eq,f}
,\end{equation}
\begin{equation}
P_{eq,f}=\frac{g_f}{3}\int dP~ p^2 f_{eq,f}
.\end{equation}
In the Navier-Stokes equation, the heat flow is related to the gradients of temperature and pressure \cite{Greif:2013bb} as
\begin{equation}\label{J}
Q^i=- \kappa\delta^{ij}\left[\partial_j T - \frac{T}{\varepsilon + P}\partial_j P\right]
.\end{equation}
Here, $\kappa$ represents the thermal conductivity. One can find the expression of the thermal conductivity by comparing eq. (\ref{K}) and eq. (\ref{J}). 

In order to obtain the expression of thermal conductivity, we start with the relativistic Boltzmann transport equation, which can be rewritten as
	\begin{equation}\label{m2}
		\begin{split}
			p^{\mu}\partial_{\mu}T\frac{\partial f_{eq,f}}{\partial T}+p^{\mu} \partial_\mu(p^{\nu}u_{\nu})\frac{\partial f_{eq,f}}{\partial p^0}+ q_f\Bigr[F^{0i}p_i\frac{\partial f_{eq,f}}{\partial p^0}+F^{i0}p_0\frac{\partial f_{eq,f}}{\partial p^i}\Bigl]
			=\textbf{C}\bigl[\textit{f}_\textit{f}\hspace{1mm}\bigr].
		\end{split}	
	\end{equation}
With the help of eq. (\ref{xy8}), the above equation can be written in the novel RTA model as
	\begin{multline}\label{mbx2}
		p^{\mu}\partial_{\mu}T\frac{\partial f_{eq,f}}{\partial T}+p^{\mu} \partial_\mu(p^{\nu}u_{\nu})\frac{\partial f_{eq,f}}{\partial p^0}+ q_f\Bigr[F^{0i}p_i\frac{\partial f_{eq,f}}{\partial p^0}+F^{i0}p_0\frac{\partial f_{eq,f}}{\partial p^i}\Bigl]\\
		=-\frac{\omega_f}{\tau_{fp}}\biggr[\delta f_f-
		\frac{\left\langle({\omega_f}/{\tau_{fp}})\delta f_f\right\rangle_0}{\left\langle{\omega_f}/{\tau_{fp}}\right\rangle_0}+
		P^{(0)}_1\frac{\left\langle({\omega_f}/{\tau_{fp}})P^{(0)}_1\delta f_f\right\rangle_0}{\left\langle({\omega_f}/{\tau_{fp}})P^{(0)}_1P^{(0)}_1\right\rangle_0}\\
		+
  p^{\left\langle\mu\right\rangle}\frac{\left\langle({\omega_f}/{\tau_{fp}})p_{\left\langle\mu\right\rangle}\delta f_f\right\rangle_0}{(1/3)\left\langle({\omega_f}/{\tau_{fp}})p_{\left\langle\mu\right\rangle}p^{\left\langle\mu\right\rangle}\right\rangle_0}\biggl],	
\end{multline}
where $p_0 = \omega_f - \mu_f $ and $p_0 \approx \omega_f$ for very small $\mu_f$. After solving the above equation one can extract the expression of $\delta f_f$ (for a detailed calculation see appendix \ref{appendix D}) as
\begin{multline}\label{mx2}	
\delta f_f=-\frac{\tau_{fp}f_{eq,f}(1-f_{eq,f})}{TA}\Bigr[\frac{p_0}{T}\partial_0 T
+\Bigr(\frac{p_0-h_f}{p_0}\Bigl)\frac{p^i}{T}\Bigr(\partial_i T-\frac{T}{\varepsilon_f+P_f}\partial_i P_f\Bigl)\\+T \partial_0\Bigr(\frac{\mu_f}{T}\Bigl)-\frac{p^i p^{\nu}}{p_0}\partial_i u_{\nu}
-\frac{2 q_f}{p_0}\textbf{E}\cdot\textbf{p}\Bigl]
.\end{multline}	
Similarly for antiquarks, we have
\begin{multline}\label{xyyy8}	
	\delta {\bar{f}}_f=-\frac{\tau_{{\bar{fp}}}{\bar{f}}_{eq,f}(1-{\bar{f}}_{eq,f})}{T\bar{A}}\Bigr[\frac{p_0}{T}\partial_0 T
	+\Bigr(\frac{p_0-\bar {h}_f}{p_0}\Bigl)\frac{p^i}{T}\Bigr(\partial_i T-\frac{T}{\varepsilon_f+P_f}\partial_i P_f\Bigl) \\ - T \partial_0\Bigr(\frac{\mu_f}{T}\Bigl)-\frac{p^i p^{\nu}}{p_0}\partial_i u_{\nu}
	-\frac{2 q_{\bar{f}}}{p_0}\textbf{E}\cdot\textbf{p}\Bigl].	
\end{multline}	
Finally, replacing the values of $\delta f_f$ and $\delta {\bar{f}}_f$ in eq. (\ref{K}) and comparing it with eq. (\ref{J}), we get the final expression of the thermal conductivity in the novel RTA model as
	\begin{equation}\label{x19}
	\kappa^{Novel-RTA}=\frac{\beta^2}{6\pi^2}\sum_{f}g_f\int dp \frac{p^4}{\omega^2_f }\Biggl[ \frac{(\omega_f-h_f)^2\tau_{fp} f_{eq,f}(1-f_{eq,f})}{A}+ \frac{(\omega_f-\bar{h}_f)^2\tau_{{\bar{fp}}} {\bar{f}}_{eq,f}(1-{\bar{f}}_{eq,f})}{\bar{A}}
	\Biggl].
\end{equation}

In the novel BGK model, the relativistic Boltzmann transport equation takes the following form, 
\begin{multline}
	p^{\mu}\partial_{\mu}T\frac{\partial f_{eq,f}}{\partial T}+p^{\mu} \partial_\mu(p^{\nu}u_{\nu})\frac{\partial f_{eq,f}}{\partial p^0}+ q_f\Bigr[F^{0i}p_i\frac{\partial f_{eq,f}}{\partial p^0}+F^{i0}p_0\frac{\partial f_{eq,f}}{\partial p^i}\Bigl]\\
 =-\frac{p^\mu u_\mu}{\tau_{fp}}\Bigg[\delta f_f-
	\frac{\left\langle({\omega_f}/{\tau_{fp}})\delta f_f\right\rangle_0}{\left\langle{\omega_f}/{\tau_{fp}}\right\rangle_0}
 +
 P^{(0)}_1\frac{\left\langle({\omega_f}/{\tau_{fp}})P^{(0)}_1\delta f_f\right\rangle_0}{\left\langle({\omega_f}/{\tau_{fp}})P^{(0)}_1P^{(0)}_1\right\rangle_0}+ p^{\left\langle\mu\right\rangle}\frac{\left\langle({\omega_f}/{\tau_{fp}})p_{\left\langle\mu\right\rangle}\delta f_f\right\rangle_0}{(1/3)\left\langle({\omega_f}/{\tau_{fp}})p_{\left\langle\mu\right\rangle}p^{\left\langle\mu\right\rangle}\right\rangle_0}\Bigg]\\
 + g_f n^{-1}_{eq,f} (p^\mu u_\mu) f_{eq,f}\int \frac{p^\mu u_\mu}{\tau_{fp}}\Bigg[ \delta f_f-
	\frac{\left\langle({\omega_f}/{\tau_{fp}})\delta f_f\right\rangle_0}{\left\langle{\omega_f}/{\tau_{fp}}\right\rangle_0} +
	P^{(0)}_1\frac{\left\langle({\omega_f}/{\tau_{fp}})P^{(0)}_1\delta f_f\right\rangle_0}{\left\langle({\omega_f}/{\tau_{fp}})P^{(0)}_1P^{(0)}_1\right\rangle_0}\\
	+ p^{\left\langle\mu\right\rangle}\frac{\left\langle({\omega_f}/{\tau_{fp}})p_{\left\langle\mu\right\rangle}\delta f_f\right\rangle_0}{(1/3)\left\langle({\omega_f}/{\tau_{fp}})p_{\left\langle\mu\right\rangle}p^{\left\langle\mu\right\rangle}\right\rangle_0}\Bigg] dP.
	\end{multline}
Now, neglecting higher order terms $\left(\mathcal O(\delta f_f)^2\right)$, we get the solution of this equation as
\begin{equation}\label{dr2}
\delta f_f=\delta f^{(0)}_f +g_f n^{-1}_{eq,f} f_{eq,f}\int_{ p\prime } \delta f^{(0)}_f
,\end{equation}
where 
\begin{multline}
	\delta f^{(0)}_f=-\frac{\tau_{fp}f_{eq,f}(1-f_{eq,f})}{TA}\Bigr[\frac{p_0}{T}\partial_0 T
	+\Bigr(\frac{p_0-h_f}{p_0}\Bigl)\frac{p^i}{T}\Bigr(\partial_i T-\frac{T}{\varepsilon_f+P_f}\partial_i P_f\Bigl)\\+T \partial_0\Bigr(\frac{\mu_f}{T}\Bigl)-\frac{p^i p^{\nu}}{p_0}\partial_i u_{\nu}
	-\frac{2 q_f}{p_0}\textbf{E}\cdot\textbf{p}\Bigl].
\end{multline}
The infinitesimal change in the quark distribution function is obtained (for a detailed calculation see appendix \ref{appendix E}) as
\begin{multline}\label{xxx51}
	\delta f_f=-\frac{\tau_{fp}f_{eq,f}(1-f_{eq,f})}{TA}\Bigr[\frac{p_0}{T}\partial_0 T
	+\Bigr(\frac{p_0-h_f}{p_0}\Bigl)\frac{p^i}{T}\Bigr(\partial_i T-\frac{T}{\varepsilon_f+P_f}\partial_i P_f\Bigl)+T \partial_0\Bigr(\frac{\mu_f}{T}\Bigl)-\frac{p^i p^{\nu}}{p_0}\partial_i u_{\nu}\\~~~~~~
	-\frac{2 q_f}{p_0}\textbf{E}\cdot\textbf{p}\Bigl]- g_f n^{-1}_{eq,f} f_{eq,f} \int_{p \prime}\frac{\tau_{fp}f_{eq,f}(1-f_{eq,f})}{TA}\Bigr[\frac{p_0}{T}\partial_0 T
	+\Bigr(\frac{p_0-h_f}{p_0}\Bigl)\frac{p^i}{T}\Bigr(\partial_i T-\frac{T}{\varepsilon_f+P_f}\partial_i P_f\Bigl)\\+T \partial_0\Bigr(\frac{\mu_f}{T}\Bigl)-\frac{p^i p^{\nu}}{p_0}\partial_i u_{\nu}
	-\frac{2 q_f}{p_0}\textbf{E}\cdot\textbf{p}\Bigl] .
\end{multline}
Similarly, $\delta {\bar{f}}_f$ for antiquark case is obtained as
\begin{multline}\label{xxxy3}
		\delta {\bar{f}}_f=-\frac{\tau_{{\bar{fp}}}{\bar{f}}_{eq,f}(1-{\bar{f}}_{eq,f})}{T\bar{A}}\Bigr[\frac{p_0}{T}\partial_0 T
	+\Bigr(\frac{p_0-\bar{h}_f}{p_0}\Bigl)\frac{p^i}{T}\Bigr(\partial_i T-\frac{T}{\varepsilon_f+P_f}\partial_i P_f\Bigl)  - T \partial_0\Bigr(\frac{\mu_f}{T}\Bigl)-\frac{p^i p^{\nu}}{p_0}\partial_i u_{\nu}\\~~~~~~
	-\frac{2 q_{\bar{f}}}{p_0}\textbf{E}\cdot\textbf{p}\Bigl]- g_f n^{-1}_{eq,f} {\bar{f}}_{eq,f} \int_{p \prime}\frac{\tau_{{\bar{fp}}}{\bar{f}}_{eq,f}(1-{\bar{f}}_{eq,f})}{T\bar{A}}\Bigr[\frac{p_0}{T}\partial_0 T
	+\Bigr(\frac{p_0-\bar{h}_f}{p_0}\Bigl)\frac{p^i}{T}\Bigr(\partial_i T-\frac{T}{\varepsilon_f+P_f}\partial_i P_f\Bigl) \\ - T \partial_0\Bigr(\frac{\mu_f}{T}\Bigl)-\frac{p^i p^{\nu}}{p_0}\partial_i u_{\nu}
	-\frac{2 q_{\bar{f}}}{p_0}\textbf{E}\cdot\textbf{p}\Bigl].
\end{multline} 
Finally, replacing the values of $\delta f_f$ and $\delta {\bar{f}}_f$ in eq. (\ref{K}) and comparing it with eq. (\ref{J}), we get the final expression of the thermal conductivity in the novel BGK model as
\begin{multline}\label{xx19}
	\kappa^{Novel-BGK}=\frac{\beta^2}{6\pi^2}\sum_{f}g_f\int dp \frac{p^4}{\omega^2_f}\Biggl[ \frac{(\omega_f-h_f)^2\tau_{fp} f_{eq,f}(1-f_{eq,f})}{A} +\frac{(\omega_f-\bar{h}_f)^2 \tau_ {\bar{fp}}\bar{f}_{eq,f}(1-\bar{f}_{eq,f})}{\bar{A}}\Biggr]\\+\frac{\beta^2}{\sqrt{3}}\sum_{f}g^2_f  n^{-1}_{eq,f}\Biggl[\int_{ p} \frac{p}{\omega_f}(\omega_f-h_f)f_{eq,f} \int_{ p\prime}\frac{p\prime}{\omega_f}   \frac{(\omega_f-h_f)\tau_{fp} f_{eq,f}(1-f_{eq,f})}{A}\\
 +\int_{ p} \frac{p}{\omega_f}(\omega_f-\bar{h}_f) \bar{f}_{eq,f} \int_{ p\prime}\frac{p\prime}{\omega_f} \frac{(\omega_f-\bar{h}_f) \tau_ {\bar{fp}}\bar{f}_{eq,f}(1-\bar{f}_{eq,f})}{\bar{A}}\Biggr].
\end{multline}

\section{Results and discussions}
The interaction of the partons with the thermal medium gives rise to a certain mass known as the thermal mass or the quasiparticle mass. Within the framework of the quasiparticle model, we may see QGP as a system of massive noninteracting quasiparticles. The quasiparticle quark-gluon plasma (qQGP) model is a commonly employed framework to explain the collective properties of the QGP medium. The quasiparticle masses of particles in a dense thermal medium rely on both temperature and chemical potential. For a dense QCD 
medium, the thermal masses (squared) of quark and gluon up to one-loop are given \cite{Braaten:1991gm, Peshier:2002ww} by
\begin{eqnarray}\label{N}
&& m_{fT}^2=\frac{g^2T^2}{6}\left(1+\frac{\mu_f^2}{\pi^2T^2}\right), \\
&&\label{G.M.}m_{gT}^2=\frac{g^2T^2}{6}\left(N_c+\frac{N_f}{2}+\frac{3}{2\pi^2T^2}\sum_f\mu_f^2\right)
,\end{eqnarray}
respectively, with $g=(4\pi\alpha_s)^{{1}/{2}}$, where the expression of $\alpha_s$ is given in eq. \eqref{ww2}. It can be seen from eq. \eqref{ww2} that, $\alpha_s$ depends on both temperature and 
chemical potential. In this work, we have used $\mu_f=\mu=0.04$ GeV, corroborating consistency 
with the theoretical predictions and experimental data \cite{Abelev:PRC79'2009,Bazavov:PRL109'2012,Roland:PPNP77'2014,arXiv:1502.02730}. The chosen value of the chemical potential is utilized to approximate realistic conditions in highly energetic heavy ion collisions, 
ensuring the numerical stability in the calculations. The accessible energy scales for a thermal QCD medium are related to the temperature, quark mass and chemical potential. The QGP phase, which is made up of light quarks, light antiquarks and gluons, has a temperature high enough to reach the maximum energy scale. The hard thermal loop (HTL) perturbation theory is one of the effective theories that is applied to tackle the divergences that arise in the computation of the QCD thermodynamic observables, the transport coefficients 
and the amplitudes in the high temperature regime. 

\begin{figure}[H]
\begin{subfigure}[h]{0.5\textwidth}
\includegraphics[width=\textwidth]{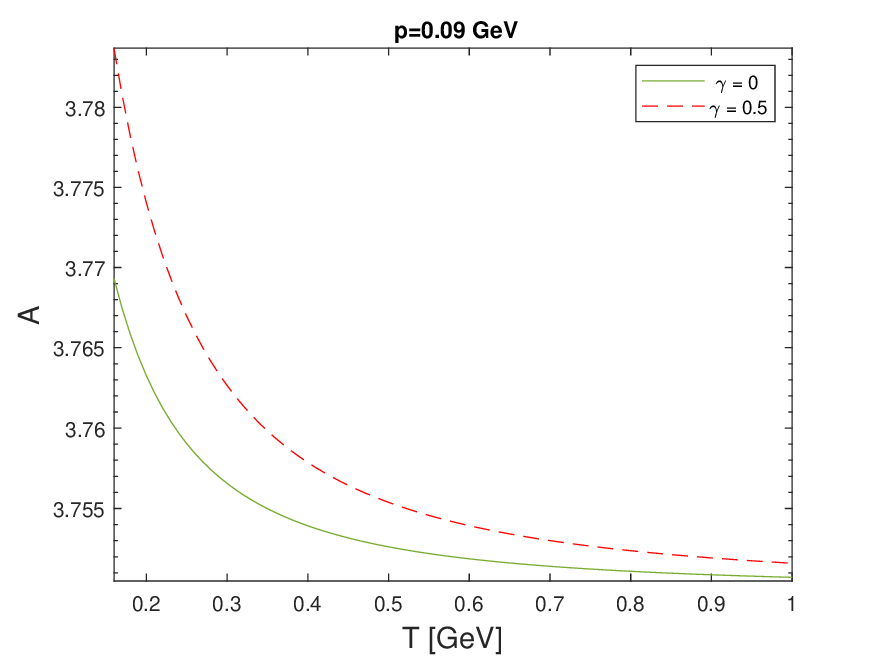}
\caption{}\label{1f1}
\end{subfigure}
\hfill
\begin{subfigure}[h]{0.5\textwidth}
\includegraphics[width=\textwidth]{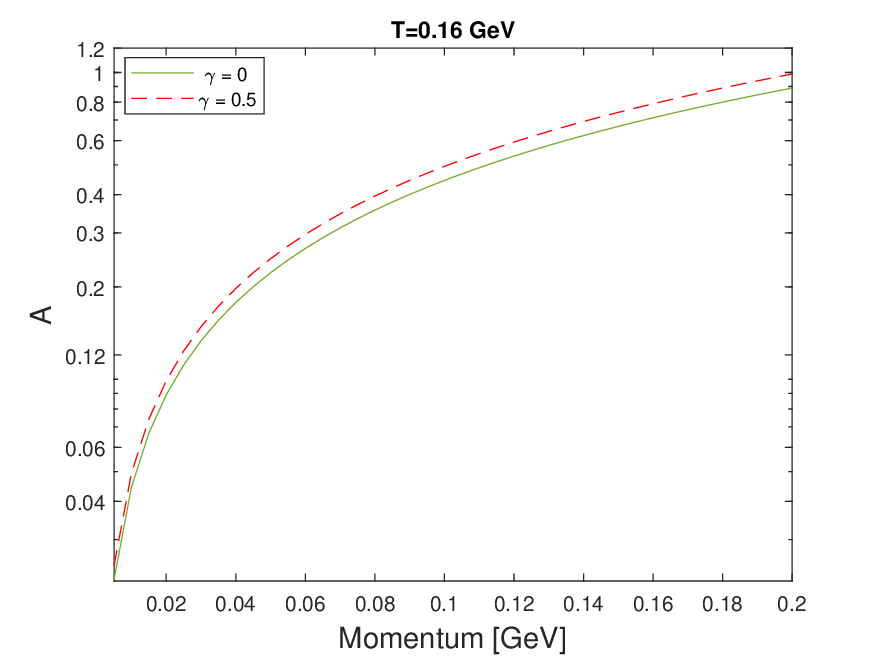}
\caption{}\label{1f2}
\end{subfigure}
\caption{Variation of A (a) as a function of temperature at a fixed momentum and (b) as a function of momentum at a fixed temperature.}\label{1f}
\end{figure}

\begin{figure}[H]
\begin{subfigure}[h]{0.5\textwidth}
\includegraphics[width=\textwidth]{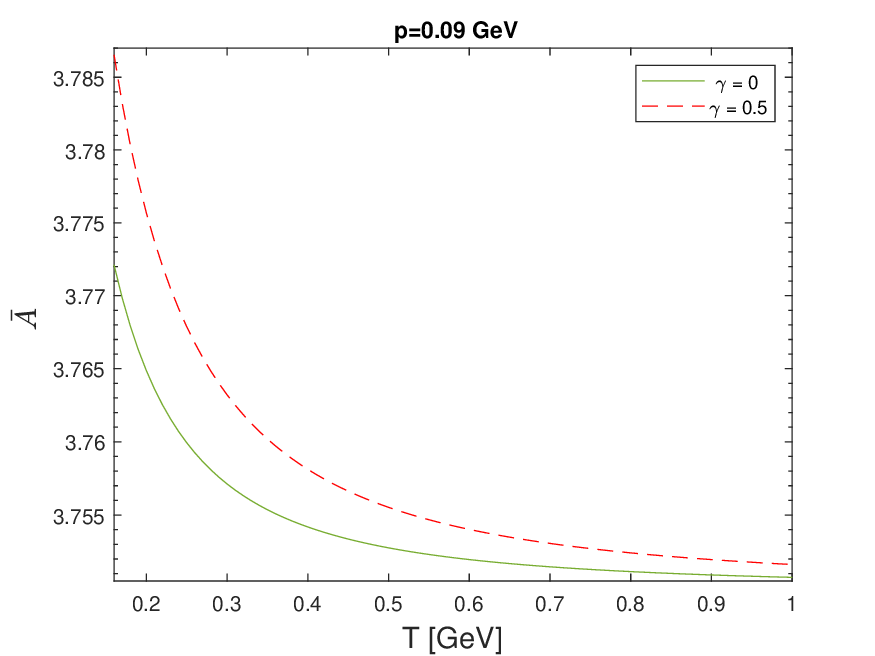}
\caption{}\label{1f1A.P.}
\end{subfigure}
\hfill
\begin{subfigure}[h]{0.5\textwidth}
\includegraphics[width=\textwidth]{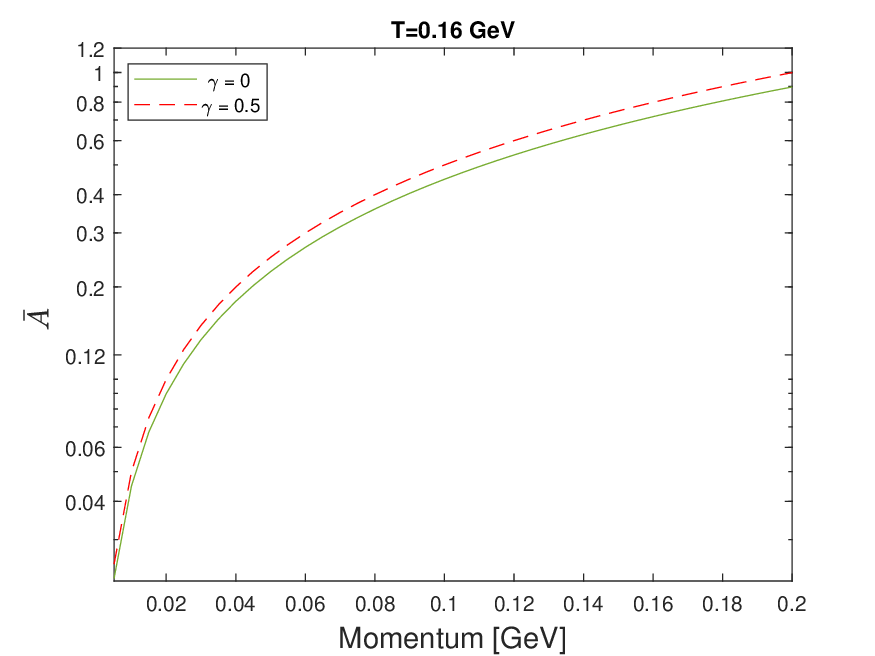}
\caption{}\label{1f2A.P.}
\end{subfigure}
\caption{Variation of $\bar{A}$ (a) as a function of temperature at a fixed momentum and (b) as a function of momentum at a fixed temperature.}\label{1fA.P.}
\end{figure}

In this section, the charge and the heat transport coefficients are investigated with the thermal masses as given in equations (\ref{N}) and (\ref{G.M.}). In the novel RTA and the novel BGK models, quantities $A$ 
and $\bar{A}$ appear in the denominators of the electrical conductivity and the thermal conductivity 
expressions. Thus the knowledge on the variations of these quantities with temperature and momentum is important to understand the variations of the conductivities. In figures \ref{1f} and \ref{1fA.P.}, we have plotted $A$ and $\bar{A}$ for the momentum-dependent relaxation time ($\gamma=0.5$) and for the momentum-independent relaxation time ($\gamma=0$), respectively. We have found decreasing trends of $A$ and $\bar{A}$ with the increase of temperature (figures \ref{1f1} and \ref{1f1A.P.}) and their increasing trends with the increase of momentum (figures \ref{1f2} and \ref{1f2A.P.}). In both the cases, it is observed that the 
momentum dependence of the relaxation time enhances both $A$ and $\bar{A}$. Thus, through the effects of the momentum-dependent relaxation time on $A$ and $\bar{A}$, it is possible to comprehend 
how the momentum-dependent relaxation time would modulate the transport coefficients. 

\begin{figure}[H]
    \begin{subfigure}[h]{0.5\textwidth}
            \includegraphics[width=\textwidth]{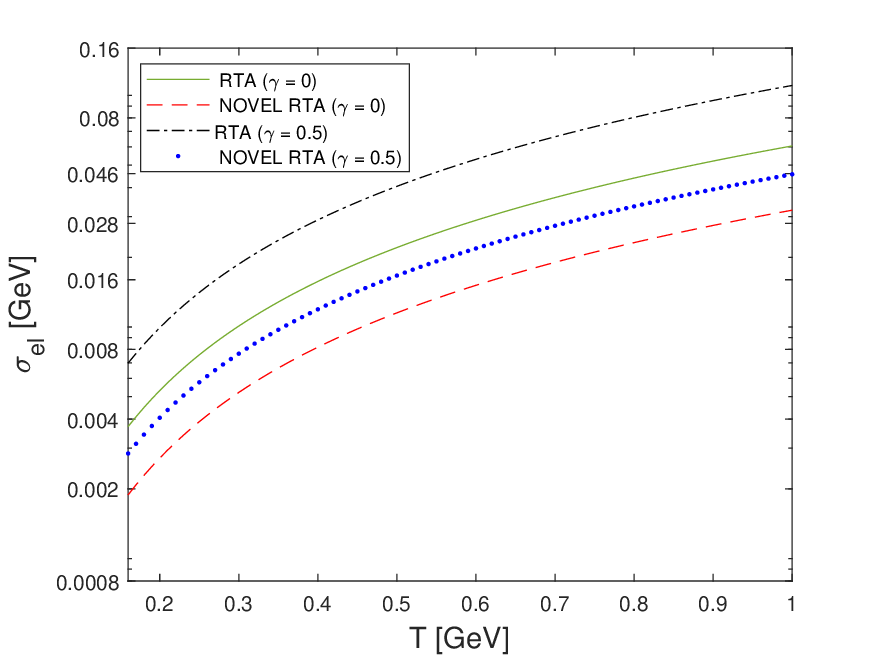}
        \caption{}\label{2f1}
    \end{subfigure}
    \begin{subfigure}[h]{0.5\textwidth}
       \includegraphics[width=\textwidth]{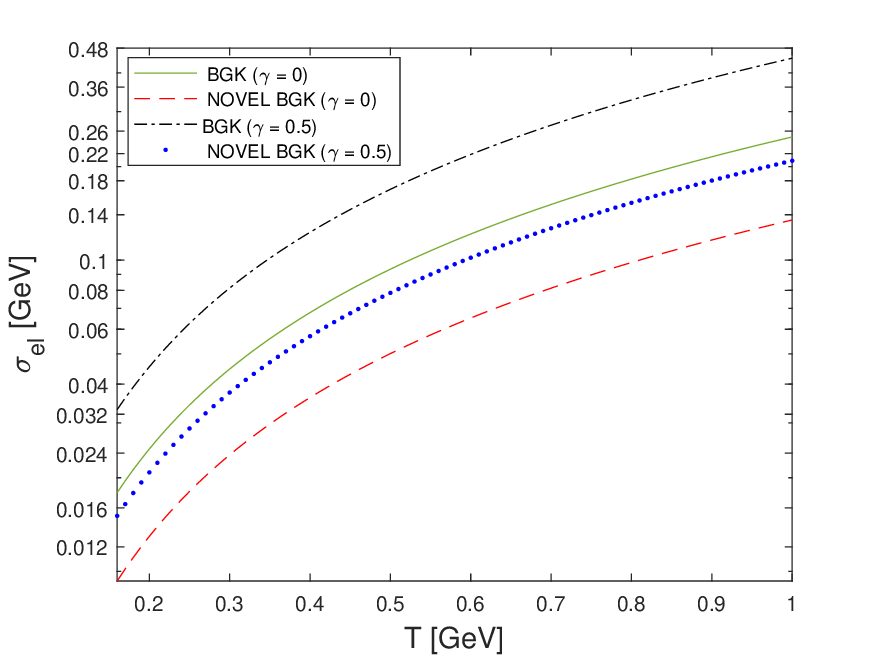}
            \caption{}\label{2f2}
    \end{subfigure}
    \caption{Variation of the electrical conductivity ($\sigma_{el}$) with the temperature, (a) comparison with the result in the RTA model and (b) comparison with the result in the BGK model.}\label{2f}
\end{figure}

Figures \ref{2f1} and \ref{2f2} show the variations of the electrical conductivity ($\sigma_{el}$) 
with the temperature in the novel RTA and the novel BGK models at finite chemical potential, 
respectively. More precisely, we have compared our results on the electrical conductivity with those of the standard RTA and the standard BGK models. We have also compared the result for $\gamma=0.5$ with that for $\gamma=0$. The relaxation time appears to be momentum-dependent (eq. \eqref{p77}) in the first case, whereas the relaxation time is momentum-independent in the second case. We have observed that, regardless of the relaxation type ($\gamma=0$ or $\gamma=0.5$), there is a noticeable reduction in the magnitudes of $\sigma_{el}$ in the novel RTA and the novel BGK models as compared to those in the RTA and the BGK models, respectively. This reduction in the electrical conductivity in the novel RTA and the novel BGK models is mainly attributed to the quantities $A$ and $\bar{A}$ appearing in the denominators of equations \eqref{oo1} and \eqref{xx3}. 

\begin{figure}[H]  
 \begin{subfigure}[h]{0.5\textwidth}
            \includegraphics[width=\textwidth]{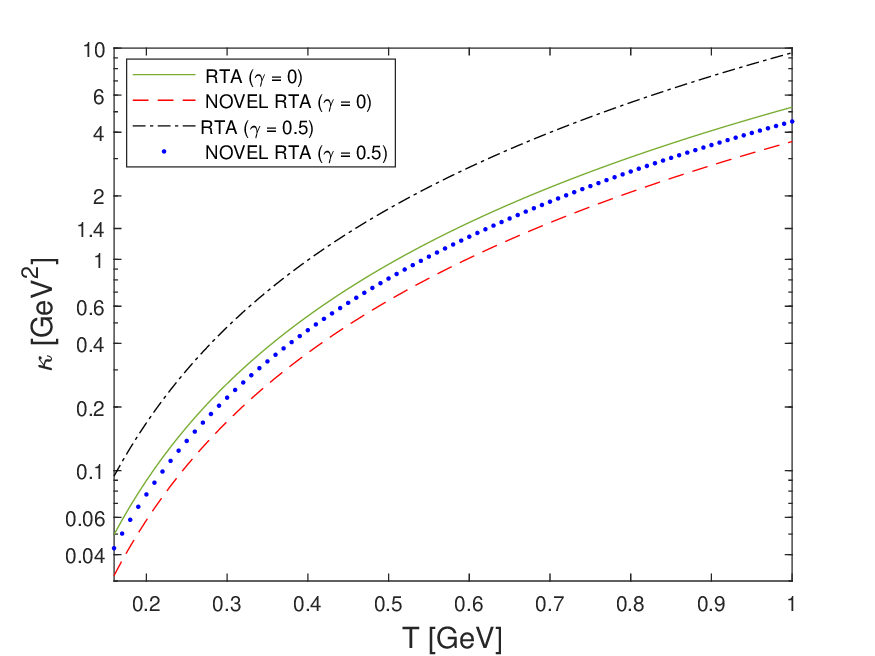}
        \caption{}\label{3f1}
    \end{subfigure}   
 \begin{subfigure}[h]{0.5\textwidth}
       
        \includegraphics[width=\textwidth]{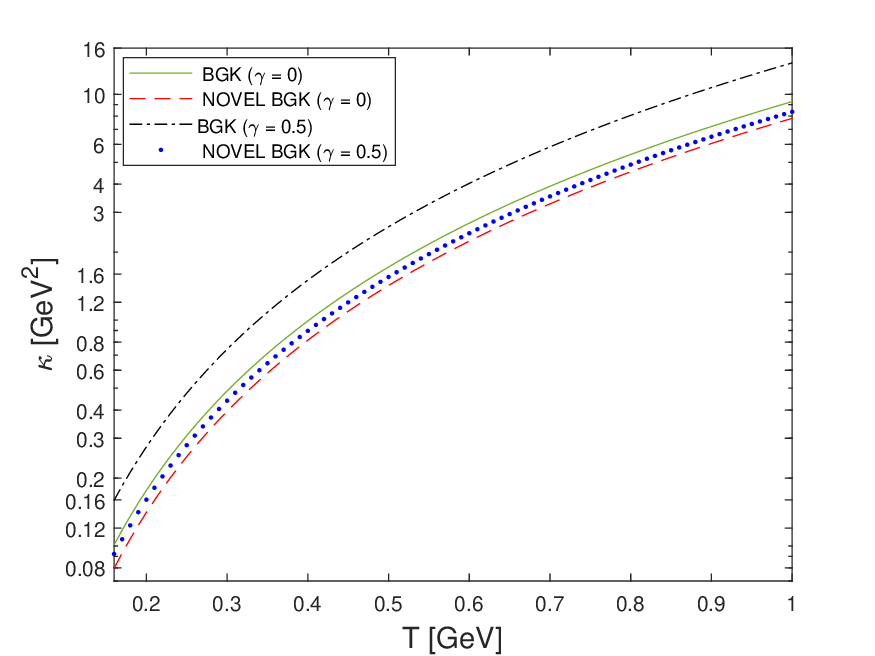}
            \caption{}\label{3f2}
    \end{subfigure}
    \caption{Variation of the thermal conductivity ($\kappa$) with the temperature, (a) comparison 
with the result in the RTA model and (b) comparison with the result in the BGK model.}\label{3f}
\end{figure}

\begin{table}[H]
    \centering
    \begin{tabular}{|c|c|c|c|c|c|c|c|c|c|} \hline  
         \multicolumn{2}{|c|}{T=0.16 GeV}&  \(\frac{\sigma_{NR}}{\sigma_{R}}\)&  \(\frac{\sigma_{NB}}{\sigma_{B}}\)&  \(\frac{\kappa_{NR}}{\kappa_{R}}\)&  \(\frac{\kappa_{NB}}{\kappa_{B}}\)&  \(\frac{D_{NR}}{D_{R}}\)&  \(\frac{D_{NB}}{D_{B}}\)&  \(\frac{L_{NR}}{L_{R}}\)& \(\frac{L_{NB}}{L_{B}}\)\\ \hline  
         \multicolumn{2}{|c|}{\(\gamma =0\)}&  0.50&  0.51&  0.64&  0.78&  0.64&  0.78&  1.26& 1.51\\ \hline  
         \multicolumn{2}{|c|}{\(\gamma = 0.5\)}&  0.40&  0.45&  0.45&  0.57&  0.45&  0.57&  1.11& 1.27\\ \hline 
    \end{tabular}
\caption{Effect of collision integral on various transport coefficients in the low temperature regime.}\label{table:1}
    \label{TABLE 1}
\end{table}
\begin{table}[H]
    \centering
    \begin{tabular}{|c|c|c|c|c|c|c|c|c|c|} \hline  
         \multicolumn{2}{|c|}{T=1 GeV}&  \(\frac{\sigma_{NR}}{\sigma_{R}}\)&  \(\frac{\sigma_{NB}}{\sigma_{B}}\)&  \(\frac{\kappa_{NR}}{\kappa_{R}}\)&  \(\frac{\kappa_{NB}}{\kappa_{B}}\)&  \(\frac{D_{NR}}{D_{R}}\)&  \(\frac{D_{NB}}{D_{B}}\)&  \(\frac{L_{NR}}{L_{R}}\)& \(\frac{L_{NB}}{L_{B}}\)\\ \hline  
         \multicolumn{2}{|c|}{\(\gamma =0\)}&  0.52&  0.54&  0.54&  0.54&  0.68&  0.83&  1.55& 1.30\\ \hline  
         \multicolumn{2}{|c|}{\(\gamma = 0.5\)}&  0.41&  0.46&  0.46&  0.46&  0.47&  0.60&  1.29& 1.26\\ \hline 
    \end{tabular}
\caption{Effect of collision integral on various transport coefficients in the high temperature regime.}\label{table:2}
    \label{TABLE 2}
\end{table}

In figures \ref{3f1} and \ref{3f2}, the thermal conductivity ($\kappa$) has been plotted as a function of the temperature in the novel RTA and the novel BGK models at finite chemical potential, respectively. These figures show the comparison of our results on the thermal conductivity with those of the standard RTA and standard BGK models. We have found that, irrespective of the relaxation type ($\gamma=0$ or $\gamma=0.5$), there is a discernible decrease in the magnitudes of $\kappa$ in the novel RTA and the novel BGK models as compared to those in the RTA and the BGK models, respectively. The quantities $A$ and $\bar{A}$ appearing in the denominators of equations \eqref{x19} and \eqref{xx19} mainly contribute in the decrease of the thermal conductivity in the novel RTA and the novel BGK models. 

The electrical and thermal conductivities are found to be significantly impacted by the adjustments we 
made to the RTA model as well as the BGK model when an energy-dependent relaxation time ($\gamma=0.5$) 
is used. Tables \ref{table:1} and \ref{table:2} show how the updated collision integral affects the 
charge and heat transfer in general in the low temperature and the high temperature regimes, respectively. Based on the electrical conductivity related ratios in tables \ref{table:1} and \ref{table:2}, we can conclude that, for the novel RTA and the novel BGK models, the drop in the magnitude of $\sigma_{el}$ becomes larger when considering the momentum dependent relaxation time as compared to the standard RTA and standard BGK models. We may infer from the thermal conductivity related ratios in tables \ref{table:1} and \ref{table:2} that, as compared to the standard RTA and standard BGK models, the reduction in the magnitude of $\kappa$ 
for the novel RTA and the novel BGK models increases when the momentum dependent relaxation time is taken into account. Thus the momentum dependence of the relaxation time decelerates the flow of charge as well as the flow of heat in the medium. 

On the whole, our study introduces a modified collision integral that improves the description of the electrical and the thermal transport properties in the QGP medium. Our updated collision integral 
considers a more accurate relaxation time dependency, which results in reduced transport coefficients, 
better matching the lattice QCD predictions. We have incorporated the momentum-dependent relaxation time ($\gamma=0.5$), capturing nontrivial scattering effects in QGP. This enables us to investigate a wider range of QCD interactions and to present a stronger suppression of charge and heat flow. Our results indicate that the QGP medium behaves as a strongly coupled fluid, exhibiting suppressed electrical and thermal conductivities in contrast to the standard RTA and standard BGK models. 

\section{Applications}
The purpose of this section is to study some applications of the electrical and the thermal conductivities 
for the hot and dense QCD medium. In subsection \ref{dc}, we analyze the rate of heat transfer in the 
medium through the thermal diffusion constant in the novel RTA and the novel BGK models. The relative behavior of the thermal conductivity and the electrical conductivity is explored in subsection \ref{ln} 
using the Lorenz number in the aforesaid models. 

\subsection{Thermal diffusion constant}\label{dc}
Instead of being measured directly, the thermal conductivity in common materials is usually determined from measurements of the thermal diffusion constant $(D_{T})$. This is related to the rate of heat transfer in the medium. The thermal diffusion constant in the first order relativistic viscous hydrodynamics is defined as 
\begin{equation}
    D_T=\frac{\kappa}{C_p}
.\end{equation}
Here, $C_p$ represents the specific heat at constant pressure and it can be determined by using the following formula. 
\begin{equation}
    C_p=\frac{\partial (\varepsilon+P)}{\partial T}
,\end{equation}
where $\varepsilon$ and $P$ are given in equations \eqref{E} and \eqref{P}, respectively. Increase 
of the thermal diffusion constant explains the faster heat transfer in the medium. 

\begin{figure}[H]
    
    \begin{subfigure}[h]{0.5\textwidth}
            \includegraphics[width=\textwidth]{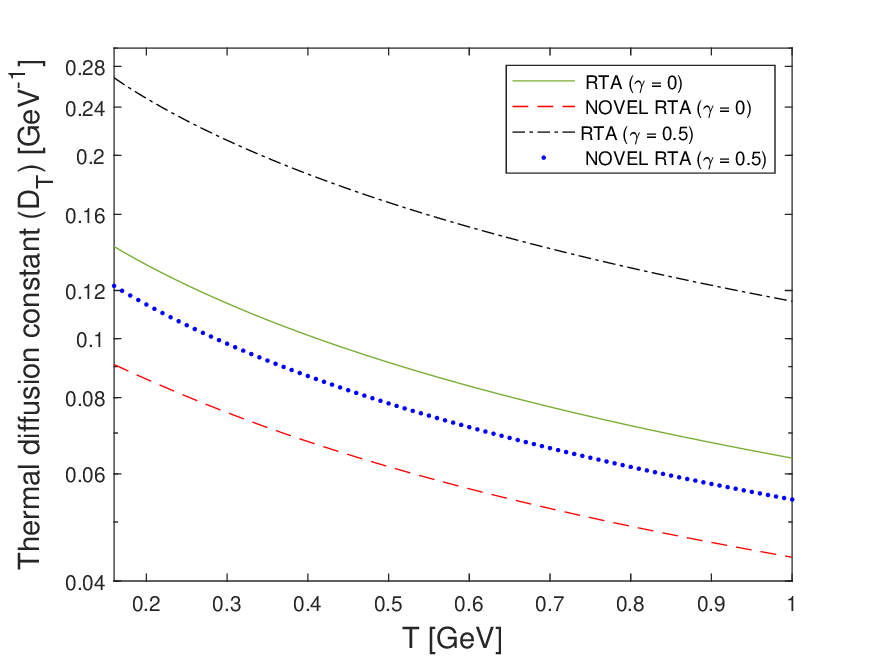}
        \caption{}\label{4f1}
    \end{subfigure}
    \begin{subfigure}[h]{0.5\textwidth}
       
        \includegraphics[width=\textwidth]{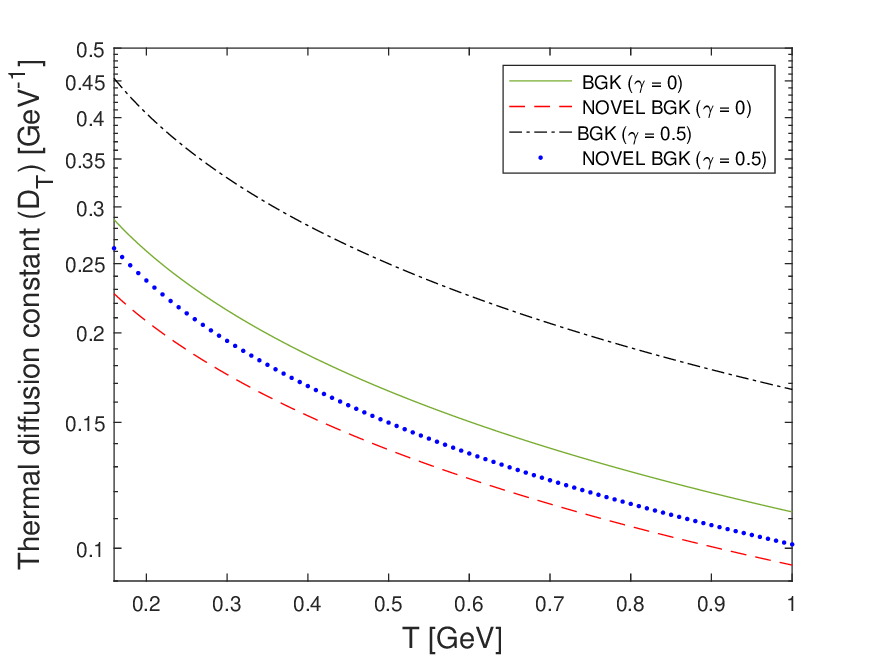}
            \caption{}\label{4f2}
    \end{subfigure}
       \caption{Variation of the thermal diffusion constant ($D_T$) with the temperature, (a) comparison with the result in the RTA model and (b) comparison with the result in the BGK model.}\label{4f}
\end{figure}

Figure \ref{4f} shows the variation of the thermal diffusion constant with the temperature at a fixed chemical potential in the novel RTA and the novel BGK models. It is evident from figures \ref{4f1} and \ref{4f2} that the magnitude of $D_{T}$ gets decreased regardless of the relaxation type ($\gamma=0$ or $\gamma=0.5$) for both RTA and BGK models with the modified integral term. When we compare it with the standard RTA and standard BGK models, we have found that the momentum-dependent relaxation time ($\gamma=0.5$) leads to a considerable 
shift in the thermal diffusion constant. We can conclude from the thermal diffusion constant related ratios 
in tables \ref{table:1} and \ref{table:2} that, for the novel RTA and the novel BGK models, the decline in the magnitude of $D_T$ becomes larger when considering the momentum-dependent relaxation time as compared to the standard RTA and standard BGK models. 

\subsection{Lorenz number}\label{ln}
The Lorenz number ($L$) is the proportionality factor in the Wiedemann-Franz law, which states that the ratio of the heat transport coefficient ($\kappa$) to the charge transport coefficient ($\sigma_{el}$) is directly proportional to the temperature, {\em i.e.}, 
\begin{equation}
\frac{\kappa}{\sigma_{el}}=LT
.\end{equation}
This law describes how the phenomena of heat transfer and charge transfer behave in relation to one another in a medium. 

\begin{figure}[H]
    \begin{subfigure}[h]{0.5\textwidth}
            \includegraphics[width=\textwidth]{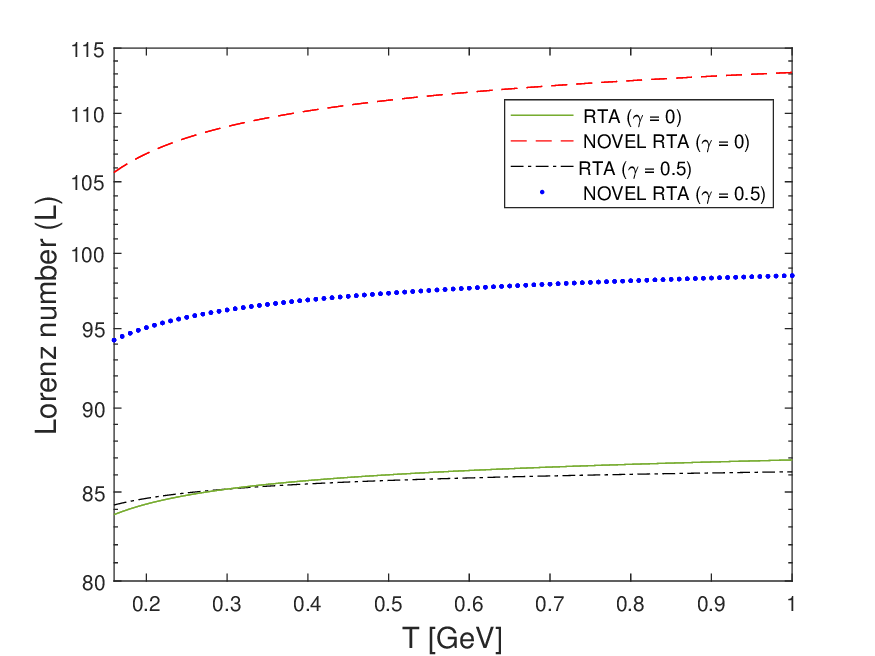}
        \caption{}\label{5f1}
    \end{subfigure}
    \begin{subfigure}[h]{0.5\textwidth}   
        \includegraphics[width=\textwidth]{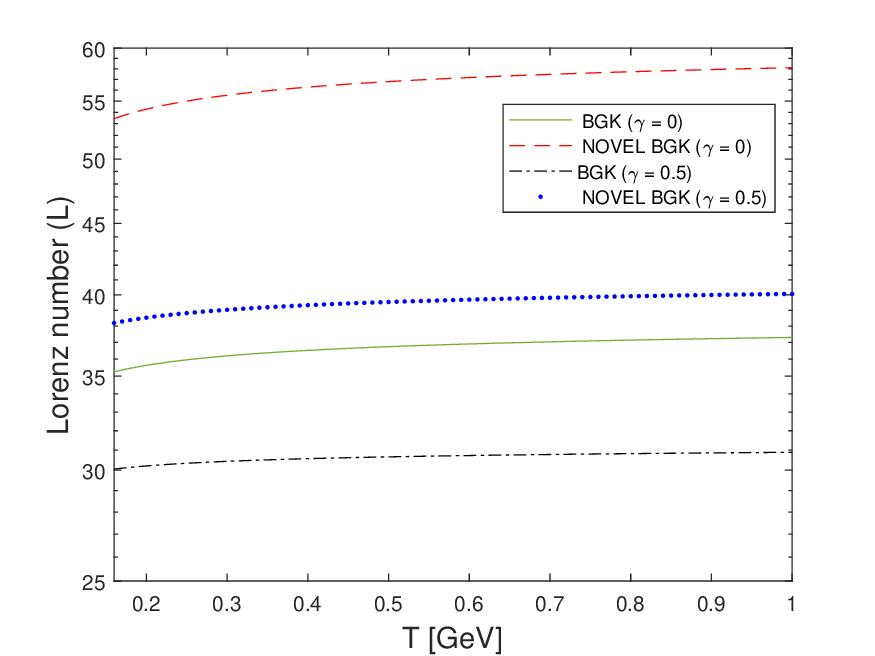}
            \caption{}\label{5f2}
    \end{subfigure}
       \caption{Variation of the Lorenz number ($L$) with the temperature, (a) comparison with the result 
in the RTA model and (b) comparison with the result in the BGK model.}\label{5f}
\end{figure}

Figure \ref{5f} shows the variation of the Lorenz number with temperature at a fixed chemical potential. It is apparent from figures \ref{5f1} and \ref{5f2} that the Lorenz number exhibits a notable increase in magnitude in case of the novel RTA and the novel BGK models, regardless of the relaxation type ($\gamma=0$ or $\gamma=0.5$). A significant change in the Lorenz number is also observed as a result of the momentum dependence of the relaxation time ($\gamma=0.5$) for both RTA and BGK models. From the Lorenz number related ratios in tables \ref{table:1} and \ref{table:2}, it is observed that, the increase in the magnitude of $L$ for the novel RTA and the novel BGK models as compared to the normal RTA and BGK models gets smaller when the momentum-dependent relaxation time is used. 

\section{Summary}
In this work, we studied the transport properties of a hot and dense QCD matter in the novel RTA and the novel BGK models. From this study, we gained more accurate understanding of the charge and heat transport phenomena in a strongly interacting thermal QCD medium with a finite chemical potential background by using the novel approaches of the aforesaid models. Using the kinetic theory technique, we solved the relativistic Boltzmann transport equation with a modified collision integral for estimating the thermal and electrical conductivities of the medium. We examined the effects of the modified collision integral on the transport of charge and heat in a thermal QCD medium, in contrast to the widely used collision terms of the RTA and the BGK models. In this research, we used the quasiparticle description of partons, where both the temperature and the chemical potential affect the constituent masses in the medium. By introducing a novel collision integral, we demonstrated that the charge and heat conductivities in QGP got significantly lower than the previously estimated results, which has direct implications for electromagnetic radiation from QGP and hydrodynamic modeling in heavy ion collisions, indicating reduced soft photon and dilepton production rates in RHIC and LHC experiments. Our findings contribute to a more accurate theoretical framework for understanding the evolution of QGP in RHIC and LHC experiments. Additionally, when we replaced the momentum-independent relaxation time ($\gamma=0$) with the momentum-dependent relaxation time ($\gamma=0.5$), we observed a discernible shift in these transport coefficients, thus explaining the impact of the momentum-dependent relaxation time on the transport properties of the medium. In the same regime, the thermal diffusion constant and the Lorenz number were also investigated. We observed that the novel collision integral leads to an increase in the Lorenz number and a decrease in the thermal diffusion constant of the medium. 

This study provides new insights into charge and heat transport phenomena in a strongly interacting QCD medium. Our findings have phenomenological implications for measurable quantities in heavy ion 
collisions. The electrical conductivity governs the interaction of QGP with electromagnetic fields, and it is related to the photon and dilepton generation from QGP via the electromagnetic current. Reduced electrical conductivity in the novel RTA and the novel BGK models as identified in our work leads to reduced photon and dilepton emission rates, hence influencing the electromagnetic signatures of the quark-gluon plasma as well as the existing experimental assessments. Reduced soft photon and dilepton yields have been seen at RHIC and LHC, suggesting that the charge transport in the quark-gluon plasma is more constrained than that predicted by standard models. Our study suggests that the QGP charge transport is more suppressed than in the traditional kinetic models, aligning with these experimental observations. Lower value of the thermal conductivity in the novel approaches of the RTA and the BGK models implies that the temperature variations persist for longer durations, leading to localized hotspots within the quark-gluon plasma. The suppression of the heat diffusion in the novel approaches of the RTA and the BGK models in our work is consistent with the strong local temperature gradients as seen at RHIC experimental facility \cite{Heinz:NPA721'2003,Shuryak:PPNP53'2004,Heinz:JPA42'2009}. Our finding regarding the increased Lorenz number in the novel RTA and the novel BGK models suggests that the QGP behaves as a strongly coupled system, where the charge transport is more suppressed than the heat transport, when compared to the predictions of the standard RTA and standard BGK models. Although our novel collision integral offers a more precise characterization of QGP transport properties, enhancements might be achieved by including the nonequilibrium corrections and higher-order moments in the Boltzmann transport equation. Subsequent research should investigate anisotropic transport coefficients in the nonequilibrium quark-gluon plasma configurations. Indirect testing of our results could be derived from experimental measures of soft photon generation, elliptic flow and jet quenching. 

\section{Acknowledgments}
One of the present authors (S. R.) acknowledges financial support from ANID Fondecyt Postdoctoral 
Grant 3240349 and S. D. acknowledges the SERB Power Fellowship, SPF/2022/000014 for the support 
on this work. 

\appendix
\appendixpage
\addappheadtotoc
\begin{appendix}
\renewcommand{\theequation}{A.\arabic{equation}}
\section{Derivation of equation (\ref{kl01})}\label{appendix A}
Using the result of ref. \cite{Rocha:2021zcw}, one can write the relativistic Boltzmann transport equation as
	\begin{multline}\label{xyz}
		q_{f}\textbf{E}\cdot\textbf{p}\frac{\partial f_{eq,f}}{\partial p^0}+q_f p_0\textbf{E}\cdot\frac{\partial f_{eq,f}}{\partial \textbf{p}}=-\frac{\omega_f}{\tau_{fp}}\biggr[\delta f_f-
		\frac{\left\langle({\omega_f}/{\tau_{fp}})\delta f_f\right\rangle_0}{\left\langle{\omega_f}/{\tau_{fp}}\right\rangle_0}\\+
		P^{(0)}_1\frac{\left\langle({\omega_f}/{\tau_{fp}})P^{(0)}_1\delta f_f\right\rangle_0}{\left\langle({\omega_f}/{\tau_{fp}})P^{(0)}_1P^{(0)}_1\right\rangle_0}
		+
  p^{\left\langle\mu\right\rangle}\frac{\left\langle({\omega_f}/{\tau_{fp}})p_{\left\langle\mu\right\rangle}\delta f_f\right\rangle_0}{(1/3)\left\langle({\omega_f}/{\tau_{fp}})p_{\left\langle\mu\right\rangle}p^{\left\langle\mu\right\rangle}\right\rangle_0}\Biggl].
	\end{multline}
The second term in the right-hand side can be reduced to 
\begin{multline}\label{xx}	
\frac{\left\langle({\omega_f}/{\tau_{fp}})\delta f_f\right\rangle_0}{\left\langle{\omega_f}/{\tau_{fp}}\right\rangle_0}
=({1}/{\left\langle{\omega_f}/{\tau_{fp}}\right\rangle_0})\int dP \frac{\omega_f}{\tau_{fp}}{ f_{eq,f}} \delta f_f
=\left( {1}/{\left\langle{\omega_f}/{\tau_{fp}}\right\rangle_0}\right) \\ \times \Biggr[ \delta f_f \int dP \frac{\omega_f}{\tau_{fp}}{ f_{eq,f}}- \int \Bigr[ \frac{d (\delta f_{f})}{dP} \int dP \frac{\omega_f}{\tau_{fp}}f_{eq,f}\Bigl]  dP\Biggl]\\=\left({1}/{\left\langle{\omega_f}/{\tau_{fp}}\right\rangle_0}\right)  \delta f_f \left({\left\langle{\omega_f}/{\tau_{fp}}\right\rangle_0}\right) 
= \delta f_f
.\end{multline}
Here, we have used the fact that the shift of the distribution function from its equilibrium value is 
infinitesimal, thus one can use the derivative $\frac{d(\delta f_{f})}{dP}\approx 0$. 
Similarly, the third and the fourth terms can be respectively reduced to 
\begin{equation}\label{yy}
	P^{(0)}_1\frac{\left\langle({\omega_f}/{\tau_{fp}})P^{(0)}_1\delta f_f\right\rangle_0}{\left\langle({\omega_f}/{\tau_{fp}})P^{(0)}_1P^{(0)}_1\right\rangle_0}=P^{(0)}_1\frac{\left\langle({\omega_f}/{\tau_{fp}})P^{(0)}_1\right\rangle_0}{\left\langle({\omega_f}/{\tau_{fp}})P^{(0)}_1P^{(0)}_1\right\rangle_0}\delta f_f,	
	\end{equation}
\begin{equation}\label{zz}
p^{\left\langle\mu\right\rangle}\frac{\left\langle({\omega_f}/{\tau_{fp}})p_{\left\langle\mu\right\rangle}\delta f_f\right\rangle_0}{(1/3)\left\langle({\omega_f}/{\tau_{fp}})p_{\left\langle\mu\right\rangle}p^{\left\langle\mu\right\rangle}\right\rangle_0}\\=p^{\left\langle\mu\right\rangle}\frac{\left\langle({\omega_f}/{\tau_{fp}})p_{\left\langle\mu\right\rangle}\right\rangle_0}{(1/3)\left\langle({\omega_f}/{\tau_{fp}})p_{\left\langle\mu\right\rangle}p^{\left\langle\mu\right\rangle}\right\rangle_0}\delta f_f	
.\end{equation}
Now, with the help of equations \eqref{xx}, \eqref{yy} and \eqref{zz}, eq. \eqref{xyz} takes the following form, 
\begin{multline}\label{xyz1}
q_{f}\textbf{E}\cdot\textbf{p}\frac{\partial f_{eq,f}}{\partial p^0}+q_f p_0\textbf{E}\cdot\frac{\partial f_{eq,f}}{\partial \textbf{p}}
	=-\frac{\omega_f}{\tau_{fp}}\biggr[
	P^{(0)}_1\frac{\left\langle({\omega_f}/{\tau_{fp}})P^{(0)}_1\right\rangle_0}{\left\langle({\omega_f}/{\tau_{fp}})P^{(0)}_1P^{(0)}_1\right\rangle_0}\delta f_f
	\\+p^{\left\langle\mu\right\rangle}\frac{\left\langle({\omega_f}/{\tau_{fp}})p_{\left\langle\mu\right\rangle}\right\rangle_0}{(1/3)\left\langle({\omega_f}/{\tau_{fp}})p_{\left\langle\mu\right\rangle}p^{\left\langle\mu\right\rangle}\right\rangle_0}\delta f_f\Biggl]=-\frac{\omega_f A}{\tau_{fp}} \delta f_f,
\end{multline}
where, 
\begin{multline}
	A=\left[1-\frac{\omega_f\int p^2 (f_{eq,f}/ \tau_{fp}) dp}{\int p^2\omega_f (f_{eq,f}/\tau_{fp})dp}\right]\frac{\int p^2 (f_{eq,f}/\tau_{fp})\Bigr[1-\Bigr(\frac{\int p^2 (f_{eq,f}/\tau_{fp})dp}{\int p^2 \omega_f (f_{eq,f}/\tau_{fp})dp}\Bigl)\omega_f\Bigl] dp}{\int p^2 (f_{eq,f}/\tau_{fp})\Bigr[1-\Bigr(\frac{\int p^2 (f_{eq,f}/\tau_{fp})dp}{\int p^2 \omega_f (f_{eq,f}/\tau_{fp})dp}\Bigl)\omega_f\Bigl]^2 dp}\\ 
	+\frac{3p\int p^3 (f_{eq,f}/ \tau_{fp}) dp}{\int p^4 (f_{eq,f}/ \tau_{fp}) dp}.
\end{multline}
Using the following partial derivatives, 
\begin{equation}\label{G}	
	\left.	\begin{aligned}
		\frac{\partial f_{eq,f}}{\partial \textbf{p}}=- \frac{\textbf{p}}{\omega_f} \beta f_{eq,f}(1-f_{eq,f}), \\
		\frac{\partial f_{eq,f}}{\partial p^0}=-\beta f_{eq,f}(1-f_{eq,f}), 
	\end{aligned}
	\right\}
\end{equation}
we solve eq. \eqref{xyz1} to get the expression of $\delta f_f$ as given in eq. \eqref{kl01}. 

\renewcommand{\theequation}{B.\arabic{equation}}
\section{The particle number and the energy-momentum conservation laws \label{appendix B}}
In this study, we have introduced a novel collision integral in the BGK model of the following form, 
\begin{multline}
\textbf{C}\bigl[\textit{f}_\textit{f}\hspace{1mm}\bigr]^{NBGK}= -\frac{p^\mu u_\mu}{\tau_{fp}}\Bigg[\delta f_f-
	\frac{\left\langle({\omega_f}/{\tau_{fp}})\delta f_f\right\rangle_0}{\left\langle{\omega_f}/{\tau_{fp}}\right\rangle_0}+
	P^{(0)}_1\frac{\left\langle({\omega_f}/{\tau_{fp}})P^{(0)}_1\delta f_f\right\rangle_0}{\left\langle({\omega_f}/{\tau_{fp}})P^{(0)}_1P^{(0)}_1\right\rangle_0}
  \\
	+ p^{\left\langle\mu\right\rangle}\frac{\left\langle({\omega_f}/{\tau_{fp}})p_{\left\langle\mu\right\rangle}\delta f_f\right\rangle_0}{(1/3)\left\langle({\omega_f}/{\tau_{fp}})p_{\left\langle\mu\right\rangle}p^{\left\langle\mu\right\rangle}\right\rangle_0}\Bigg]
 + g_f n^{-1}_{eq,f}(p^\mu u_\mu) f_{eq,f}\int \frac{p^\mu u_\mu}{\tau_{fp}}\Bigg[ \delta f_f-
	\frac{\left\langle({\omega_f}/{\tau_{fp}})\delta f_f\right\rangle_0}{\left\langle{\omega_f}/{\tau_{fp}}\right\rangle_0}\\
 +
	P^{(0)}_1\frac{\left\langle({\omega_f}/{\tau_{fp}})P^{(0)}_1\delta f_f\right\rangle_0}{\left\langle({\omega_f}/{\tau_{fp}})P^{(0)}_1P^{(0)}_1\right\rangle_0}
	+ p^{\left\langle\mu\right\rangle}\frac{\left\langle({\omega_f}/{\tau_{fp}})p_{\left\langle\mu\right\rangle}\delta f_f\right\rangle_0}{(1/3)\left\langle({\omega_f}/{\tau_{fp}})p_{\left\langle\mu\right\rangle}p^{\left\langle\mu\right\rangle}\right\rangle_0}\Bigg] dP.
\end{multline}
We will be subject to the following constraints if we insist that the conservation laws be upheld. 
\begin{equation}
    \partial_\mu N^{\mu}=\int \textbf{C}\bigl[\textit{f}_\textit{f}\hspace{1mm}\bigr]^{NBGK} dP=0,
\end{equation}

\begin{equation}
    \partial_{\mu} T^{\mu\nu}=\int p^{\nu}\textbf{C}\bigl[\textit{f}_\textit{f}\hspace{1mm}\bigr]^{NBGK} dP=0.
\end{equation}

Validity of the particle number conservation law in the novel BGK model: 
\begin{multline}\label{xyze}
    \partial_\mu N^{\mu}=\int \textbf{C}\bigl[\textit{f}_\textit{f}\hspace{1mm}\bigr]^{NBGK} dP=- \int \frac{p^\mu u_\mu}{\tau_{fp}}\Bigg[\delta f_f-
	\frac{\left\langle({\omega_f}/{\tau_{fp}})\delta f_f\right\rangle_0}{\left\langle{\omega_f}/{\tau_{fp}}\right\rangle_0}+
	P^{(0)}_1\frac{\left\langle({\omega_f}/{\tau_{fp}})P^{(0)}_1\delta f_f\right\rangle_0}{\left\langle({\omega_f}/{\tau_{fp}})P^{(0)}_1P^{(0)}_1\right\rangle_0}
  \\
	+ p^{\left\langle\mu\right\rangle}\frac{\left\langle({\omega_f}/{\tau_{fp}})p_{\left\langle\mu\right\rangle}\delta f_f\right\rangle_0}{(1/3)\left\langle({\omega_f}/{\tau_{fp}})p_{\left\langle\mu\right\rangle}p^{\left\langle\mu\right\rangle}\right\rangle_0}\Bigg] dP +\int \Bigg[ g_f n^{-1}_{eq,f} (p^\mu u_\mu) f_{eq,f}\int \frac{p^\mu u_\mu}{\tau_{fp}}\Bigg[ \delta f_f-
	\frac{\left\langle({\omega_f}/{\tau_{fp}})\delta f_f\right\rangle_0}{\left\langle{\omega_f}/{\tau_{fp}}\right\rangle_0}\\
 +
	P^{(0)}_1\frac{\left\langle({\omega_f}/{\tau_{fp}})P^{(0)}_1\delta f_f\right\rangle_0}{\left\langle({\omega_f}/{\tau_{fp}})P^{(0)}_1P^{(0)}_1\right\rangle_0}
	+ p^{\left\langle\mu\right\rangle}\frac{\left\langle({\omega_f}/{\tau_{fp}})p_{\left\langle\mu\right\rangle}\delta f_f\right\rangle_0}{(1/3)\left\langle({\omega_f}/{\tau_{fp}})p_{\left\langle\mu\right\rangle}p^{\left\langle\mu\right\rangle}\right\rangle_0}\Bigg]dP\Bigg]dP\\=
 - \int \frac{p^\mu u_\mu}{\tau_{fp}}\Bigg[\delta f_f-
	\frac{\left\langle({\omega_f}/{\tau_{fp}})\delta f_f\right\rangle_0}{\left\langle{\omega_f}/{\tau_{fp}}\right\rangle_0}+
	P^{(0)}_1\frac{\left\langle({\omega_f}/{\tau_{fp}})P^{(0)}_1\delta f_f\right\rangle_0}{\left\langle({\omega_f}/{\tau_{fp}})P^{(0)}_1P^{(0)}_1\right\rangle_0} + p^{\left\langle\mu\right\rangle}\frac{\left\langle({\omega_f}/{\tau_{fp}})p_{\left\langle\mu\right\rangle}\delta f_f\right\rangle_0}{(1/3)\left\langle({\omega_f}/{\tau_{fp}})p_{\left\langle\mu\right\rangle}p^{\left\langle\mu\right\rangle}\right\rangle_0}\Bigg] dP\\
 + \int \frac{p^\mu u_\mu}{\tau_{fp}}\Bigg[\delta f_f-
	\frac{\left\langle({\omega_f}/{\tau_{fp}})\delta f_f\right\rangle_0}{\left\langle{\omega_f}/{\tau_{fp}}\right\rangle_0}+
	P^{(0)}_1\frac{\left\langle({\omega_f}/{\tau_{fp}})P^{(0)}_1\delta f_f\right\rangle_0}{\left\langle({\omega_f}/{\tau_{fp}})P^{(0)}_1P^{(0)}_1\right\rangle_0}+ p^{\left\langle\mu\right\rangle}\frac{\left\langle({\omega_f}/{\tau_{fp}})p_{\left\langle\mu\right\rangle}\delta f_f\right\rangle_0}{(1/3)\left\langle({\omega_f}/{\tau_{fp}})p_{\left\langle\mu\right\rangle}p^{\left\langle\mu\right\rangle}\right\rangle_0}\Bigg] dP\\ \times \int g_f n^{-1}_{eq,f} (p^\mu u_\mu) f_{eq,f} dP - \int \Bigg[ \frac{d }{dP} \Bigg[\int \frac{p^\mu u_\mu}{\tau_{fp}}\Big[\delta f_f-
	\frac{\left\langle({\omega_f}/{\tau_{fp}})\delta f_f\right\rangle_0}{\left\langle{\omega_f}/{\tau_{fp}}\right\rangle_0}+
	P^{(0)}_1\frac{\left\langle({\omega_f}/{\tau_{fp}})P^{(0)}_1\delta f_f\right\rangle_0}{\left\langle({\omega_f}/{\tau_{fp}})P^{(0)}_1P^{(0)}_1\right\rangle_0}\\+ p^{\left\langle\mu\right\rangle}\frac{\left\langle({\omega_f}/{\tau_{fp}})p_{\left\langle\mu\right\rangle}\delta f_f\right\rangle_0}{(1/3)\left\langle({\omega_f}/{\tau_{fp}})p_{\left\langle\mu\right\rangle}p^{\left\langle\mu\right\rangle}\right\rangle_0}\Big]dP\Bigg]\int g_f n^{-1}_{eq,f} (p^\mu u_\mu) f_{eq,f}dP\Bigg] dP = T1 +T2 + T3 
,\end{multline}
where, 
\begin{multline}\label{xxe}
  T1= - \int \frac{p^\mu u_\mu}{\tau_{fp}}\Bigg[\delta f_f-
	\frac{\left\langle({\omega_f}/{\tau_{fp}})\delta f_f\right\rangle_0}{\left\langle{\omega_f}/{\tau_{fp}}\right\rangle_0}+
	P^{(0)}_1\frac{\left\langle({\omega_f}/{\tau_{fp}})P^{(0)}_1\delta f_f\right\rangle_0}{\left\langle({\omega_f}/{\tau_{fp}})P^{(0)}_1P^{(0)}_1\right\rangle_0} \\+ p^{\left\langle\mu\right\rangle}\frac{\left\langle({\omega_f}/{\tau_{fp}})p_{\left\langle\mu\right\rangle}\delta f_f\right\rangle_0}{(1/3)\left\langle({\omega_f}/{\tau_{fp}})p_{\left\langle\mu\right\rangle}p^{\left\langle\mu\right\rangle}\right\rangle_0}\Bigg] dP
,\end{multline}
 \begin{multline}\label{yye}
  T2= \int \frac{p^\mu u_\mu}{\tau_{fp}}\Bigg[\delta f_f-
	\frac{\left\langle({\omega_f}/{\tau_{fp}})\delta f_f\right\rangle_0}{\left\langle{\omega_f}/{\tau_{fp}}\right\rangle_0}+
	P^{(0)}_1\frac{\left\langle({\omega_f}/{\tau_{fp}})P^{(0)}_1\delta f_f\right\rangle_0}{\left\langle({\omega_f}/{\tau_{fp}})P^{(0)}_1P^{(0)}_1\right\rangle_0}\\+ p^{\left\langle\mu\right\rangle}\frac{\left\langle({\omega_f}/{\tau_{fp}})p_{\left\langle\mu\right\rangle}\delta f_f\right\rangle_0}{(1/3)\left\langle({\omega_f}/{\tau_{fp}})p_{\left\langle\mu\right\rangle}p^{\left\langle\mu\right\rangle}\right\rangle_0}\Bigg] dP \times \int g_f n^{-1}_{eq,f} (p^\mu u_\mu) f_{eq,f} dP \\
 = \int \frac{p^\mu u_\mu}{\tau_{fp}}\Bigg[\delta f_f-
	\frac{\left\langle({\omega_f}/{\tau_{fp}})\delta f_f\right\rangle_0}{\left\langle{\omega_f}/{\tau_{fp}}\right\rangle_0}+
	P^{(0)}_1\frac{\left\langle({\omega_f}/{\tau_{fp}})P^{(0)}_1\delta f_f\right\rangle_0}{\left\langle({\omega_f}/{\tau_{fp}})P^{(0)}_1P^{(0)}_1\right\rangle_0}~~~~~~~~~~~~~~~~~~~~~~~~~~~~\\+ p^{\left\langle\mu\right\rangle}\frac{\left\langle({\omega_f}/{\tau_{fp}})p_{\left\langle\mu\right\rangle}\delta f_f\right\rangle_0}{(1/3)\left\langle({\omega_f}/{\tau_{fp}})p_{\left\langle\mu\right\rangle}p^{\left\langle\mu\right\rangle}\right\rangle_0}\Bigg] dP\times n^{-1}_{eq,f} \Big(g_f\int (p^\mu u_\mu) f_{eq,f} dP\Big) \\
 =\int \frac{p^\mu u_\mu}{\tau_{fp}}\Bigg[\delta f_f-
	\frac{\left\langle({\omega_f}/{\tau_{fp}})\delta f_f\right\rangle_0}{\left\langle{\omega_f}/{\tau_{fp}}\right\rangle_0}+
	P^{(0)}_1\frac{\left\langle({\omega_f}/{\tau_{fp}})P^{(0)}_1\delta f_f\right\rangle_0}{\left\langle({\omega_f}/{\tau_{fp}})P^{(0)}_1P^{(0)}_1\right\rangle_0}~~~~~~~~~~~~~~~~~~~~~~~~~~~~\\+ p^{\left\langle\mu\right\rangle}\frac{\left\langle({\omega_f}/{\tau_{fp}})p_{\left\langle\mu\right\rangle}\delta f_f\right\rangle_0}{(1/3)\left\langle({\omega_f}/{\tau_{fp}})p_{\left\langle\mu\right\rangle}p^{\left\langle\mu\right\rangle}\right\rangle_0}\Bigg] dP\times n^{-1}_{eq,f} \Big(n_{eq,f}\Big)~~~~~~~\Big[{\rm Using ~ eq.} ~ \eqref{ck} \Big]\\
 =\int \frac{p^\mu u_\mu}{\tau_{fp}}\Bigg[\delta f_f-
	\frac{\left\langle({\omega_f}/{\tau_{fp}})\delta f_f\right\rangle_0}{\left\langle{\omega_f}/{\tau_{fp}}\right\rangle_0}+
	P^{(0)}_1\frac{\left\langle({\omega_f}/{\tau_{fp}})P^{(0)}_1\delta f_f\right\rangle_0}{\left\langle({\omega_f}/{\tau_{fp}})P^{(0)}_1P^{(0)}_1\right\rangle_0}~~~~~~~~~~~~~~~~~~~~~~~~~~~~\\+ p^{\left\langle\mu\right\rangle}\frac{\left\langle({\omega_f}/{\tau_{fp}})p_{\left\langle\mu\right\rangle}\delta f_f\right\rangle_0}{(1/3)\left\langle({\omega_f}/{\tau_{fp}})p_{\left\langle\mu\right\rangle}p^{\left\langle\mu\right\rangle}\right\rangle_0}\Bigg] dP
,\end{multline}
\begin{multline}\label{zze}
 T3=- \int \Bigg[ \frac{d }{dP}  \Bigg[\int \frac{p^\mu u_\mu}{\tau_{fp}}\Big[\delta f_f-
	\frac{\left\langle({\omega_f}/{\tau_{fp}})\delta f_f\right\rangle_0}{\left\langle{\omega_f}/{\tau_{fp}}\right\rangle_0}+
	P^{(0)}_1\frac{\left\langle({\omega_f}/{\tau_{fp}})P^{(0)}_1\delta f_f\right\rangle_0}{\left\langle({\omega_f}/{\tau_{fp}})P^{(0)}_1P^{(0)}_1\right\rangle_0}\\+ p^{\left\langle\mu\right\rangle}\frac{\left\langle({\omega_f}/{\tau_{fp}})p_{\left\langle\mu\right\rangle}\delta f_f\right\rangle_0}{(1/3)\left\langle({\omega_f}/{\tau_{fp}})p_{\left\langle\mu\right\rangle}p^{\left\langle\mu\right\rangle}\right\rangle_0}\Big]dP\Bigg] \int g_f n^{-1}_{eq,f} (p^\mu u_\mu) f_{eq,f}dP\Bigg] dP \\
 =0.~~~~~~~~~~~~~~~~~~~~~~~~~~~~~~~~~~~~~~~~~~~~~~~~~~~~~~~~~~~~~~~~\Big[{\rm Due ~ to ~ \mathcal O((\delta f_f)^2)}\Big]
\end{multline}
Now, putting the results of equations \eqref{xxe}, \eqref{yye} and \eqref{zze} in eq. \eqref{xyze}, we get 
\begin{multline}\label{x1x}
    \partial_\mu N^{\mu}=- \int \frac{p^\mu u_\mu}{\tau_{fp}}\Bigg[\delta f_f-
	\frac{\left\langle({\omega_f}/{\tau_{fp}})\delta f_f\right\rangle_0}{\left\langle{\omega_f}/{\tau_{fp}}\right\rangle_0}+
	P^{(0)}_1\frac{\left\langle({\omega_f}/{\tau_{fp}})P^{(0)}_1\delta f_f\right\rangle_0}{\left\langle({\omega_f}/{\tau_{fp}})P^{(0)}_1P^{(0)}_1\right\rangle_0} \\+ p^{\left\langle\mu\right\rangle}\frac{\left\langle({\omega_f}/{\tau_{fp}})p_{\left\langle\mu\right\rangle}\delta f_f\right\rangle_0}{(1/3)\left\langle({\omega_f}/{\tau_{fp}})p_{\left\langle\mu\right\rangle}p^{\left\langle\mu\right\rangle}\right\rangle_0}\Bigg] dP + \int \frac{p^\mu u_\mu}{\tau_{fp}}\Bigg[\delta f_f-
	\frac{\left\langle({\omega_f}/{\tau_{fp}})\delta f_f\right\rangle_0}{\left\langle{\omega_f}/{\tau_{fp}}\right\rangle_0}+
	P^{(0)}_1\frac{\left\langle({\omega_f}/{\tau_{fp}})P^{(0)}_1\delta f_f\right\rangle_0}{\left\langle({\omega_f}/{\tau_{fp}})P^{(0)}_1P^{(0)}_1\right\rangle_0} \\+ p^{\left\langle\mu\right\rangle}\frac{\left\langle({\omega_f}/{\tau_{fp}})p_{\left\langle\mu\right\rangle}\delta f_f\right\rangle_0}{(1/3)\left\langle({\omega_f}/{\tau_{fp}})p_{\left\langle\mu\right\rangle}p^{\left\langle\mu\right\rangle}\right\rangle_0}\Bigg] dP\\
 =0.~~~~~~~~~~~~~~~~~~~~~~~~~~~~~~~~~~~~~~~~~~~~~~~~~~~~~~~~~~~~~~~~~~~~~~~~~~~~~~~~~~~~~~~~~
\end{multline}
From eq. \eqref{x1x}, it is confirmed that the particle number is instantaneously conserved in the novel BGK model. 

Validity of the energy-momentum conservation law in the novel BGK model: 
\begin{multline}
    \partial_{\mu} T^{\mu\nu}=\int p^{\nu}\textbf{C}\bigl[\textit{f}_\textit{f}\hspace{1mm}\bigr]^{NBGK} dP=- \int p^{\nu} \frac{p^\mu u_\mu}{\tau_{fp}}\Bigg[\delta f_f-
	\frac{\left\langle({\omega_f}/{\tau_{fp}})\delta f_f\right\rangle_0}{\left\langle{\omega_f}/{\tau_{fp}}\right\rangle_0}+
	P^{(0)}_1\frac{\left\langle({\omega_f}/{\tau_{fp}})P^{(0)}_1\delta f_f\right\rangle_0}{\left\langle({\omega_f}/{\tau_{fp}})P^{(0)}_1P^{(0)}_1\right\rangle_0}
  \\
	+ p^{\left\langle\mu\right\rangle}\frac{\left\langle({\omega_f}/{\tau_{fp}})p_{\left\langle\mu\right\rangle}\delta f_f\right\rangle_0}{(1/3)\left\langle({\omega_f}/{\tau_{fp}})p_{\left\langle\mu\right\rangle}p^{\left\langle\mu\right\rangle}\right\rangle_0}\Bigg] dP +\int \Bigg[ p^{\nu} g_f n^{-1}_{eq,f} (p^\mu u_\mu) f_{eq,f}\int \frac{p^\mu u_\mu}{\tau_{fp}}\Bigg[ \delta f_f-
	\frac{\left\langle({\omega_f}/{\tau_{fp}})\delta f_f\right\rangle_0}{\left\langle{\omega_f}/{\tau_{fp}}\right\rangle_0}\\
 +
	P^{(0)}_1\frac{\left\langle({\omega_f}/{\tau_{fp}})P^{(0)}_1\delta f_f\right\rangle_0}{\left\langle({\omega_f}/{\tau_{fp}})P^{(0)}_1P^{(0)}_1\right\rangle_0}
	+ p^{\left\langle\mu\right\rangle}\frac{\left\langle({\omega_f}/{\tau_{fp}})p_{\left\langle\mu\right\rangle}\delta f_f\right\rangle_0}{(1/3)\left\langle({\omega_f}/{\tau_{fp}})p_{\left\langle\mu\right\rangle}p^{\left\langle\mu\right\rangle}\right\rangle_0}\Bigg]dP\Bigg]dP\\=
 - \int  p^{\nu} \frac{p^\mu u_\mu}{\tau_{fp}}\Bigg[\delta f_f-
	\frac{\left\langle({\omega_f}/{\tau_{fp}})\delta f_f\right\rangle_0}{\left\langle{\omega_f}/{\tau_{fp}}\right\rangle_0}+
	P^{(0)}_1\frac{\left\langle({\omega_f}/{\tau_{fp}})P^{(0)}_1\delta f_f\right\rangle_0}{\left\langle({\omega_f}/{\tau_{fp}})P^{(0)}_1P^{(0)}_1\right\rangle_0} + p^{\left\langle\mu\right\rangle}\frac{\left\langle({\omega_f}/{\tau_{fp}})p_{\left\langle\mu\right\rangle}\delta f_f\right\rangle_0}{(1/3)\left\langle({\omega_f}/{\tau_{fp}})p_{\left\langle\mu\right\rangle}p^{\left\langle\mu\right\rangle}\right\rangle_0}\Bigg] dP\\
 + \int \frac{p^\mu u_\mu}{\tau_{fp}}\Bigg[\delta f_f-
	\frac{\left\langle({\omega_f}/{\tau_{fp}})\delta f_f\right\rangle_0}{\left\langle{\omega_f}/{\tau_{fp}}\right\rangle_0}+
	P^{(0)}_1\frac{\left\langle({\omega_f}/{\tau_{fp}})P^{(0)}_1\delta f_f\right\rangle_0}{\left\langle({\omega_f}/{\tau_{fp}})P^{(0)}_1P^{(0)}_1\right\rangle_0}+ p^{\left\langle\mu\right\rangle}\frac{\left\langle({\omega_f}/{\tau_{fp}})p_{\left\langle\mu\right\rangle}\delta f_f\right\rangle_0}{(1/3)\left\langle({\omega_f}/{\tau_{fp}})p_{\left\langle\mu\right\rangle}p^{\left\langle\mu\right\rangle}\right\rangle_0}\Bigg] dP\\ \times \int  p^{\nu} g_f n^{-1}_{eq,f} (p^\mu u_\mu) f_{eq,f} dP - \int \Bigg[ \frac{d }{dP}  \Bigg[\int \frac{p^\mu u_\mu}{\tau_{fp}}\Big[\delta f_f-
	\frac{\left\langle({\omega_f}/{\tau_{fp}})\delta f_f\right\rangle_0}{\left\langle{\omega_f}/{\tau_{fp}}\right\rangle_0}+
	P^{(0)}_1\frac{\left\langle({\omega_f}/{\tau_{fp}})P^{(0)}_1\delta f_f\right\rangle_0}{\left\langle({\omega_f}/{\tau_{fp}})P^{(0)}_1P^{(0)}_1\right\rangle_0}\\+ p^{\left\langle\mu\right\rangle}\frac{\left\langle({\omega_f}/{\tau_{fp}})p_{\left\langle\mu\right\rangle}\delta f_f\right\rangle_0}{(1/3)\left\langle({\omega_f}/{\tau_{fp}})p_{\left\langle\mu\right\rangle}p^{\left\langle\mu\right\rangle}\right\rangle_0}\Big]dP\Bigg] \int  p^{\nu} g_f n^{-1}_{eq,f} (p^\mu u_\mu) f_{eq,f}dP\Bigg] dP\\
  =S1^\nu+S2^\nu+S3^\nu ,~~~~~~~~~~~~~~~~~~~~~~~~~~~~~~~~~~~~~~~~~~~~~~~~~~~~~~~~~~~~~~~~~~~~~~~~~~~~~~~~~~~~~~~~~~~~~~~~
\end{multline}
where $S1^\nu$, $S2^\nu$ and $S3^\nu$ are respectively expressed as
\begin{multline}
S1^\nu= - \int p^{\nu} \frac{p^\mu u_\mu}{\tau_{fp}}\Bigg[\delta f_f-
	\frac{\left\langle({\omega_f}/{\tau_{fp}})\delta f_f\right\rangle_0}{\left\langle{\omega_f}/{\tau_{fp}}\right\rangle_0}+
	P^{(0)}_1\frac{\left\langle({\omega_f}/{\tau_{fp}})P^{(0)}_1\delta f_f\right\rangle_0}{\left\langle({\omega_f}/{\tau_{fp}})P^{(0)}_1P^{(0)}_1\right\rangle_0} \\+ p^{\left\langle\mu\right\rangle}\frac{\left\langle({\omega_f}/{\tau_{fp}})p_{\left\langle\mu\right\rangle}\delta f_f\right\rangle_0}{(1/3)\left\langle({\omega_f}/{\tau_{fp}})p_{\left\langle\mu\right\rangle}p^{\left\langle\mu\right\rangle}\right\rangle_0}\Bigg] dP \\=
  - \int p^{\nu} \frac{p^\mu u_\mu}{\tau_{fp}}\delta f_f  dP + \int p^{\nu} \frac{p^\mu u_\mu}{\tau_{fp}}\frac{\left\langle({\omega_f}/{\tau_{fp}})\delta f_f\right\rangle_0}{\left\langle{\omega_f}/{\tau_{fp}}\right\rangle_0} dP~~~~~~~~~~~~~~~~~~~~~~~~~~~~~~~\\
  - \int p^{\nu} \frac{p^\mu u_\mu}{\tau_{fp}}P^{(0)}_1\frac{\left\langle({\omega_f}/{\tau_{fp}})P^{(0)}_1\delta f_f\right\rangle_0}{\left\langle({\omega_f}/{\tau_{fp}})P^{(0)}_1P^{(0)}_1\right\rangle_0} dP\\~~~~~~~~~~~~~~~~~~~~~~
  -\int p^{\nu} \frac{p^\mu u_\mu}{\tau_{fp}}p^{\left\langle\mu\right\rangle}\frac{\left\langle({\omega_f}/{\tau_{fp}})p_{\left\langle\mu\right\rangle}\delta f_f\right\rangle_0}{(1/3)\left\langle({\omega_f}/{\tau_{fp}})p_{\left\langle\mu\right\rangle}p^{\left\langle\mu\right\rangle}\right\rangle_0} dP \\
  =S11^\nu+S12^\nu+S13^\nu+S14^\nu ,~~~~~~~~~~~~~~~~~~~~~~~~~~~~~~~~~~~~~~~~~~~~~~~~~~~~~~~~~~~~~~
\end{multline}
\begin{multline}
S2^\nu=\int \frac{p^\mu u_\mu}{\tau_{fp}}\Bigg[\delta f_f-
	\frac{\left\langle({\omega_f}/{\tau_{fp}})\delta f_f\right\rangle_0}{\left\langle{\omega_f}/{\tau_{fp}}\right\rangle_0}+
	P^{(0)}_1\frac{\left\langle({\omega_f}/{\tau_{fp}})P^{(0)}_1\delta f_f\right\rangle_0}{\left\langle({\omega_f}/{\tau_{fp}})P^{(0)}_1P^{(0)}_1\right\rangle_0}\\ 
 ~~~~~~~~~~~~~~~~~~~+ p^{\left\langle\mu\right\rangle}\frac{\left\langle({\omega_f}/{\tau_{fp}})p_{\left\langle\mu\right\rangle}\delta f_f\right\rangle_0}{(1/3)\left\langle({\omega_f}/{\tau_{fp}})p_{\left\langle\mu\right\rangle}p^{\left\langle\mu\right\rangle}\right\rangle_0}\Bigg] dP \times \int  p^{\nu} g_f n^{-1}_{eq,f} (p^\mu u_\mu) f_{eq,f} dP\\
 = \Bigg[ \int  \frac{p^\mu u_\mu}{\tau_{fp}}\delta f_f  dP - \int  \frac{p^\mu u_\mu}{\tau_{fp}}\frac{\left\langle({\omega_f}/{\tau_{fp}})\delta f_f\right\rangle_0}{\left\langle{\omega_f}/{\tau_{fp}}\right\rangle_0} dP~~~~~~~~~~~~~~~~~~~~~~~~~~~~~~~\\
  + \int  \frac{p^\mu u_\mu}{\tau_{fp}}P^{(0)}_1\frac{\left\langle({\omega_f}/{\tau_{fp}})P^{(0)}_1\delta f_f\right\rangle_0}{\left\langle({\omega_f}/{\tau_{fp}})P^{(0)}_1P^{(0)}_1\right\rangle_0} dP\\~~~~~~~~~~~~~~~~~~~~~~
  +\int  \frac{p^\mu u_\mu}{\tau_{fp}}p^{\left\langle\mu\right\rangle}\frac{\left\langle({\omega_f}/{\tau_{fp}})p_{\left\langle\mu\right\rangle}\delta f_f\right\rangle_0}{(1/3)\left\langle({\omega_f}/{\tau_{fp}})p_{\left\langle\mu\right\rangle}p^{\left\langle\mu\right\rangle}\right\rangle_0} dP\Bigg]\int  p^{\nu} g_f n^{-1}_{eq,f} (p^\mu u_\mu) f_{eq,f} dP \\
  =(S21+S22+S23+S24)S20^\nu ,~~~~~~~~~~~~~~~~~~~~~~~~~~~~~~~~~~~~~~~~~~~~~~~~~~~~~~~~~~~~~~
\end{multline}
\begin{multline}
S3^\nu=-\int \Bigg[ \frac{d }{dP}  \Bigg[\int \frac{p^\mu u_\mu}{\tau_{fp}}\Big[\delta f_f-
	\frac{\left\langle({\omega_f}/{\tau_{fp}})\delta f_f\right\rangle_0}{\left\langle{\omega_f}/{\tau_{fp}}\right\rangle_0}+
	P^{(0)}_1\frac{\left\langle({\omega_f}/{\tau_{fp}})P^{(0)}_1\delta f_f\right\rangle_0}{\left\langle({\omega_f}/{\tau_{fp}})P^{(0)}_1P^{(0)}_1\right\rangle_0}\\+ p^{\left\langle\mu\right\rangle}\frac{\left\langle({\omega_f}/{\tau_{fp}})p_{\left\langle\mu\right\rangle}\delta f_f\right\rangle_0}{(1/3)\left\langle({\omega_f}/{\tau_{fp}})p_{\left\langle\mu\right\rangle}p^{\left\langle\mu\right\rangle}\right\rangle_0}\Big]dP\Bigg] \int p^{\nu} g_f n^{-1}_{eq,f} (p^\mu u_\mu) f_{eq,f} dP\Bigg] dP
.\end{multline}
The energy-momentum conservation requires that, 
\begin{eqnarray}\label{xyzee}
&&\nonumber u_{\nu} \partial_{\mu} T^{\mu\nu}=0 \\ && \implies u_{\nu}\left(S1^\nu+S2^\nu+S3^\nu\right)=0
.\end{eqnarray}
The term $u_{\nu}S1^\nu$ can be calculated as follows. 
\begin{eqnarray}
u_{\nu}S1^\nu=u_{\nu}\left(S11^\nu+S12^\nu+S13^\nu+S14^\nu\right)
,\end{eqnarray}
where, 
\begin{multline}
u_\nu S11^\nu=  - u_\nu\int p^{\nu} \frac{p^\mu u_\mu}{\tau_{fp}}\delta f_f  dP,~~~~~~~~~~~~~~~~~~~~~~~  
\end{multline}
\begin{multline}
 u_\nu S12^\nu= u_\nu\int p^{\nu} \frac{p^\mu u_\mu}{\tau_{fp}} \frac{\left\langle({\omega_f}/{\tau_{fp}})\delta f_f\right\rangle_0}{\left\langle{\omega_f}/{\tau_{fp}}\right\rangle_0} dP=  u_\nu\int p^{\nu} \frac{p^\mu u_\mu}{\tau_{fp}} \frac{\left\langle({\omega_f}/{\tau_{fp}})\right\rangle_0}{\left\langle{\omega_f}/{\tau_{fp}}\right\rangle_0}\delta f_f dP\\
 = u_\nu\int p^{\nu} \frac{p^\mu u_\mu}{\tau_{fp}}\delta f_f  dP,~~~~~~~~~~~~~~~~\Big[ {\rm Using ~ eq.} ~ \eqref{xx} \Big]
\end{multline}
\begin{multline}
 u_\nu S13^\nu=  - u_\nu  \int p^{\nu} \frac{p^\mu u_\mu}{\tau_{fp}}P^{(0)}_1\frac{\left\langle({\omega_f}/{\tau_{fp}})P^{(0)}_1\delta f_f\right\rangle_0}{\left\langle({\omega_f}/{\tau_{fp}})P^{(0)}_1P^{(0)}_1\right\rangle_0} dP \\
 =  - u_\nu  \int p^{\nu} \frac{p^\mu u_\mu}{\tau_{fp}}P^{(0)}_1\frac{\left\langle({\omega_f}/{\tau_{fp}})P^{(0)}_1\right\rangle_0}{\left\langle({\omega_f}/{\tau_{fp}})P^{(0)}_1P^{(0)}_1\right\rangle_0} \delta f_f dP ~~~~~~~~~~~~~~\Big[{\rm Using ~ eq.} ~ \eqref{yy} \Big]\\
 =- u_\nu \frac{\left\langle({\omega_f}/{\tau_{fp}})P^{(0)}_1\right\rangle_0}{\left\langle({\omega_f}/{\tau_{fp}})P^{(0)}_1P^{(0)}_1\right\rangle_0} \int p^{\nu} \frac{p^\mu u_\mu}{\tau_{fp}}P^{(0)}_1 \delta f_f dP \\
 +  u_\nu  \int \Bigg[ \frac{d }{dP}  \Bigg[\frac{\left\langle({\omega_f}/{\tau_{fp}})P^{(0)}_1\right\rangle_0}{\left\langle({\omega_f}/{\tau_{fp}})P^{(0)}_1P^{(0)}_1\right\rangle_0}\Bigg] \int p^{\nu} \frac{p^\mu u_\mu}{\tau_{fp}}P^{(0)}_1 \delta f_f dP\Bigg]dP \\
 = - u_\nu \frac{\left\langle({\omega_f}/{\tau_{fp}})P^{(0)}_1\right\rangle_0}{\left\langle({\omega_f}/{\tau_{fp}})P^{(0)}_1P^{(0)}_1\right\rangle_0} \int p^{\nu} \frac{p^\mu u_\mu}{\tau_{fp}}P^{(0)}_1 \delta f_f dP,~~~~~~~\Big[{\rm Second ~ term ~ vanishes ~ due ~ to ~ \mathcal O((\delta f_f)^2)}\Big]
\end{multline}
\begin{multline}
 u_\nu S14^\nu=  - u_\nu  \int p^{\nu} \frac{p^\mu u_\mu}{\tau_{fp}}p^{\left\langle\mu\right\rangle}\frac{\left\langle({\omega_f}/{\tau_{fp}})p_{\left\langle\mu\right\rangle}\delta f_f\right\rangle_0}{(1/3)\left\langle({\omega_f}/{\tau_{fp}})p_{\left\langle\mu\right\rangle}p^{\left\langle\mu\right\rangle}\right\rangle_0} dP \\
 =- u_\nu  \int p^{\nu} \frac{p^\mu u_\mu}{\tau_{fp}}p^{\left\langle\mu\right\rangle}\frac{\left\langle({\omega_f}/{\tau_{fp}})p_{\left\langle\mu\right\rangle}\right\rangle_0}{(1/3)\left\langle({\omega_f}/{\tau_{fp}})p_{\left\langle\mu\right\rangle}p^{\left\langle\mu\right\rangle}\right\rangle_0}  \delta f_fdP ~~~~~~~~~~~~~~\Big[{\rm Using ~ eq.} ~ \eqref{zz} \Big]\\
 =- u_\nu \frac{\left\langle({\omega_f}/{\tau_{fp}})p_{\left\langle\mu\right\rangle}\right\rangle_0}{(1/3)\left\langle({\omega_f}/{\tau_{fp}})p_{\left\langle\mu\right\rangle}p^{\left\langle\mu\right\rangle}\right\rangle_0} \int p^{\nu} \frac{p^\mu u_\mu}{\tau_{fp}}p^{\left\langle\mu\right\rangle}  \delta f_fdP \\
 + u_\nu  \int \Bigg[ \frac{d }{dP} \Bigg[\frac{\left\langle({\omega_f}/{\tau_{fp}})p_{\left\langle\mu\right\rangle}\right\rangle_0}{(1/3)\left\langle({\omega_f}/{\tau_{fp}})p_{\left\langle\mu\right\rangle}p^{\left\langle\mu\right\rangle}\right\rangle_0}\Bigg] \int p^{\nu} \frac{p^\mu u_\mu}{\tau_{fp}}p^{\left\langle\mu\right\rangle}  \delta f_fdP\Bigg]dP \\
 = - u_\nu \frac{\left\langle({\omega_f}/{\tau_{fp}})p_{\left\langle\mu\right\rangle}\right\rangle_0}{(1/3)\left\langle({\omega_f}/{\tau_{fp}})p_{\left\langle\mu\right\rangle}p^{\left\langle\mu\right\rangle}\right\rangle_0} \int p^{\nu} \frac{p^\mu u_\mu}{\tau_{fp}}p^{\left\langle\mu\right\rangle}  \delta f_fdP.~~~~~\Big[{\rm Second ~ term ~ vanishes ~ due ~ to ~ \mathcal O((\delta f_f)^2)}\Big]
\end{multline}
Now, the term $u_{\nu}S1^\nu$ turns out to be
\begin{multline}\label{xxee}
u_\nu S1^\nu= -\Bigg[ u_\nu \frac{\left\langle({\omega_f}/{\tau_{fp}})P^{(0)}_1\right\rangle_0}{\left\langle({\omega_f}/{\tau_{fp}})P^{(0)}_1P^{(0)}_1\right\rangle_0} \int p^{\nu} \frac{p^\mu u_\mu}{\tau_{fp}}P^{(0)}_1 \delta f_f dP\\~~~~~~~~~~~~~~~~~~~~~~~~~~~~~~~~~~~~~~~~~
  + u_\nu \frac{\left\langle({\omega_f}/{\tau_{fp}})p_{\left\langle\mu\right\rangle}\right\rangle_0}{(1/3)\left\langle({\omega_f}/{\tau_{fp}})p_{\left\langle\mu\right\rangle}p^{\left\langle\mu\right\rangle}\right\rangle_0} \int p^{\nu} \frac{p^\mu u_\mu}{\tau_{fp}}p^{\left\langle\mu\right\rangle}  \delta f_fdP\Bigg] \\
  = -\Bigg[ u_\nu \frac{\left\langle({\omega_f}/{\tau_{fp}})P^{(0)}_1\right\rangle_0}{\left\langle({\omega_f}/{\tau_{fp}})P^{(0)}_1P^{(0)}_1\right\rangle_0} \int p^{\nu} \frac{p^\mu u_\mu}{\tau_{fp}}P^{(0)}_1 \phi_f f_{eq,f} dP\\~~~~~~~~~~~~~~~~~~~~~~~~~~~~~~~~~~~~~~
  + u_\nu \frac{\left\langle({\omega_f}/{\tau_{fp}})p_{\left\langle\mu\right\rangle}\right\rangle_0}{(1/3)\left\langle({\omega_f}/{\tau_{fp}})p_{\left\langle\mu\right\rangle}p^{\left\langle\mu\right\rangle}\right\rangle_0} \int p^{\nu} \frac{p^\mu u_\mu}{\tau_{fp}}p^{\left\langle\mu\right\rangle}  \phi_f f_{eq,f} dP\Bigg],
  \\~~~~~~~~~~~~~~~
  =-\Bigg[  \frac{\left\langle({\omega_f}/{\tau_{fp}})P^{(0)}_1\right\rangle_0}{\left\langle({\omega_f}/{\tau_{fp}})P^{(0)}_1P^{(0)}_1\right\rangle_0} \left\langle \omega_f P^{(0)}_1 \phi_f\right\rangle +  \frac{\left\langle({\omega_f}/{\tau_{fp}})p_{\left\langle\mu\right\rangle}\right\rangle_0}{(1/3)\left\langle({\omega_f}/{\tau_{fp}})p_{\left\langle\mu\right\rangle}p^{\left\langle\mu\right\rangle}\right\rangle_0}\left\langle \omega_f p^{\left\langle\mu\right\rangle} \phi_f\right\rangle  \Bigg]
.\end{multline}
In getting the above equation, we have used the following substitutions (in this case, $f_f$ is expressed in terms of an equilibrium contribution $(f_{eq,f})$ and a nonequilibrium correction ($\phi_f$) as
$f_f\equiv (1+\phi_f) ~f_{eq,f}$), 
\begin{eqnarray}
&&\delta f_f=\phi_f f_{eq,f}, \\ 
&&\left\langle \omega_f P^{(0)}_1\phi_f\right\rangle=\int \left({\omega_f}/{\tau_{fp}}\right) \omega_f P^{(0)}_1 \phi_f f_{eq,f}dP, \\ 
&&\left\langle \omega_f p^{\left\langle\mu\right\rangle}\phi_f\right\rangle=\int \left({\omega_f}/{\tau_{fp}}\right) \omega_f p^{\left\langle\mu\right\rangle} \phi_f f_{eq,f}dP
.\end{eqnarray}
The term $u_{\nu}S2^\nu$ can be calculated as follows. 
\begin{eqnarray}
u_{\nu}S2^\nu=u_{\nu}\left(S21+S22+S23+S24\right)S20^\nu
,\end{eqnarray}
where, 
\begin{eqnarray}
\nonumber u_{\nu}S20^\nu &=& u_{\nu} \int  p^{\nu} (p^\mu u_\mu) g_f n^{-1}_{eq,f} f_{eq,f} dP \\ &=& n^{-1}_{eq,f}g_f u_{\nu}\int  p^{\nu} (p^\mu u_\mu)  f_{eq,f} dP=\frac{\varepsilon_{eq,f}}{n_{eq,f}},~~~~~\Big[{\rm Using ~ eq.} ~ \eqref{bb} \Big]
\end{eqnarray}
\begin{multline}
  S21= \int  \frac{p^\mu u_\mu}{\tau_{fp}}\delta f_f  dP,~~~~~~~~~~~~~~~~~~~~~~~  
\end{multline}
\begin{multline}
 S22=- \int  \frac{p^\mu u_\mu}{\tau_{fp}} \frac{\left\langle({\omega_f}/{\tau_{fp}})\delta f_f\right\rangle_0}{\left\langle{\omega_f}/{\tau_{fp}}\right\rangle_0} dP= - \int \frac{p^\mu u_\mu}{\tau_{fp}} \frac{\left\langle({\omega_f}/{\tau_{fp}})\right\rangle_0}{\left\langle{\omega_f}/{\tau_{fp}}\right\rangle_0}\delta f_f dP\\
 = - \int  \frac{p^\mu u_\mu}{\tau_{fp}}\delta f_f  dP,~~~~~~~~~~~~~~~~\Big[{\rm Using ~ eq.} ~ \eqref{xx} \Big]
\end{multline}
\begin{multline}
  S23=  \int  \frac{p^\mu u_\mu}{\tau_{fp}}P^{(0)}_1\frac{\left\langle({\omega_f}/{\tau_{fp}})P^{(0)}_1\delta f_f\right\rangle_0}{\left\langle({\omega_f}/{\tau_{fp}})P^{(0)}_1P^{(0)}_1\right\rangle_0} dP \\
 =  \int  \frac{p^\mu u_\mu}{\tau_{fp}}P^{(0)}_1\frac{\left\langle({\omega_f}/{\tau_{fp}})P^{(0)}_1\right\rangle_0}{\left\langle({\omega_f}/{\tau_{fp}})P^{(0)}_1P^{(0)}_1\right\rangle_0} \delta f_f dP ~~~~~~~~~~~~~~\Big[{\rm Using ~ eq.} ~ \eqref{yy} \Big]\\
 =  \frac{\left\langle({\omega_f}/{\tau_{fp}})P^{(0)}_1\right\rangle_0}{\left\langle({\omega_f}/{\tau_{fp}})P^{(0)}_1P^{(0)}_1\right\rangle_0} \int  \frac{p^\mu u_\mu}{\tau_{fp}}P^{(0)}_1 \delta f_f dP \\
 -    \int \Bigg[ \frac{d }{dP}  \Bigg[\frac{\left\langle({\omega_f}/{\tau_{fp}})P^{(0)}_1\right\rangle_0}{\left\langle({\omega_f}/{\tau_{fp}})P^{(0)}_1P^{(0)}_1\right\rangle_0}\Bigg] \int \frac{p^\mu u_\mu}{\tau_{fp}}P^{(0)}_1 \delta f_f dP\Bigg]dP \\
 =   \frac{\left\langle({\omega_f}/{\tau_{fp}})P^{(0)}_1\right\rangle_0}{\left\langle({\omega_f}/{\tau_{fp}})P^{(0)}_1P^{(0)}_1\right\rangle_0} \int  \frac{p^\mu u_\mu}{\tau_{fp}}P^{(0)}_1 \delta f_f dP,~~~~~~~\Big[{\rm Second ~ term ~ vanishes ~ due ~ to ~ \mathcal O((\delta f_f)^2)}\Big]
\end{multline}
\begin{multline}
  S24=  \int  \frac{p^\mu u_\mu}{\tau_{fp}}p^{\left\langle\mu\right\rangle}\frac{\left\langle({\omega_f}/{\tau_{fp}})p_{\left\langle\mu\right\rangle}\delta f_f\right\rangle_0}{(1/3)\left\langle({\omega_f}/{\tau_{fp}})p_{\left\langle\mu\right\rangle}p^{\left\langle\mu\right\rangle}\right\rangle_0} dP \\
 =  \int  \frac{p^\mu u_\mu}{\tau_{fp}}p^{\left\langle\mu\right\rangle}\frac{\left\langle({\omega_f}/{\tau_{fp}})p_{\left\langle\mu\right\rangle}\right\rangle_0}{(1/3)\left\langle({\omega_f}/{\tau_{fp}})p_{\left\langle\mu\right\rangle}p^{\left\langle\mu\right\rangle}\right\rangle_0}  \delta f_fdP ~~~~~~~~~~~~~~\Big[{\rm Using ~ eq.} ~ \eqref{zz}\Big]\\
 =  \frac{\left\langle({\omega_f}/{\tau_{fp}})p_{\left\langle\mu\right\rangle}\right\rangle_0}{(1/3)\left\langle({\omega_f}/{\tau_{fp}})p_{\left\langle\mu\right\rangle}p^{\left\langle\mu\right\rangle}\right\rangle_0} \int  \frac{p^\mu u_\mu}{\tau_{fp}}p^{\left\langle\mu\right\rangle}  \delta f_fdP \\
 -  \int \Bigg[ \frac{d }{dP} \Bigg[\frac{\left\langle({\omega_f}/{\tau_{fp}})p_{\left\langle\mu\right\rangle}\right\rangle_0}{(1/3)\left\langle({\omega_f}/{\tau_{fp}})p_{\left\langle\mu\right\rangle}p^{\left\langle\mu\right\rangle}\right\rangle_0}\Bigg] \int  \frac{p^\mu u_\mu}{\tau_{fp}}p^{\left\langle\mu\right\rangle}  \delta f_fdP\Bigg]dP \\
 =   \frac{\left\langle({\omega_f}/{\tau_{fp}})p_{\left\langle\mu\right\rangle}\right\rangle_0}{(1/3)\left\langle({\omega_f}/{\tau_{fp}})p_{\left\langle\mu\right\rangle}p^{\left\langle\mu\right\rangle}\right\rangle_0} \int  \frac{p^\mu u_\mu}{\tau_{fp}}p^{\left\langle\mu\right\rangle}  \delta f_fdP.~~~~~\Big[{\rm Second ~ term ~ vanishes ~ due ~ to ~ \mathcal O((\delta f_f)^2)}\Big]
\end{multline}
Thus the term $u_{\nu}S2^\nu$ becomes
\begin{multline}\label{yyee}
u_{\nu} S2^\nu= \Bigg[ u_\nu \frac{\left\langle({\omega_f}/{\tau_{fp}})P^{(0)}_1\right\rangle_0}{\left\langle({\omega_f}/{\tau_{fp}})P^{(0)}_1P^{(0)}_1\right\rangle_0} \int  \frac{p^\mu u_\mu}{\tau_{fp}}P^{(0)}_1 \delta f_f dP \\~~~~~~~~~~~~~~~~~~~~~~~~~~~~~~~~~~~~~~
  + u_\nu \frac{\left\langle({\omega_f}/{\tau_{fp}})p_{\left\langle\mu\right\rangle}\right\rangle_0}{(1/3)\left\langle({\omega_f}/{\tau_{fp}})p_{\left\langle\mu\right\rangle}p^{\left\langle\mu\right\rangle}\right\rangle_0} \int  \frac{p^\mu u_\mu}{\tau_{fp}}p^{\left\langle\mu\right\rangle}  \delta f_fdP\Bigg]\frac{\epsilon_{eq,f}}{n_{eq,f}}\\=\Bigg[ u_\nu \frac{\left\langle({\omega_f}/{\tau_{fp}})P^{(0)}_1\right\rangle_0}{\left\langle({\omega_f}/{\tau_{fp}})P^{(0)}_1P^{(0)}_1\right\rangle_0} \int  \frac{p^\mu u_\mu}{\tau_{fp}}P^{(0)}_1 \phi_f f_{eq,f} dP \\~~~~~~~~~~~~~~~~~~~~~~~~~~~~~~~~~~
  + u_\nu \frac{\left\langle({\omega_f}/{\tau_{fp}})p_{\left\langle\mu\right\rangle}\right\rangle_0}{(1/3)\left\langle({\omega_f}/{\tau_{fp}})p_{\left\langle\mu\right\rangle}p^{\left\langle\mu\right\rangle}\right\rangle_0} \int  \frac{p^\mu u_\mu}{\tau_{fp}}p^{\left\langle\mu\right\rangle}  \phi_f f_{eq,f} dP\Bigg]\frac{\epsilon_{eq,f}}{n_{eq,f}}\\~~~~~~~~~~~~~~~~
  =\Bigg[  \frac{\left\langle({\omega_f}/{\tau_{fp}})P^{(0)}_1\right\rangle_0}{\left\langle({\omega_f}/{\tau_{fp}})P^{(0)}_1P^{(0)}_1\right\rangle_0} \left\langle P^{(0)}_1 \phi_f\right\rangle +  \frac{\left\langle({\omega_f}/{\tau_{fp}})p_{\left\langle\mu\right\rangle}\right\rangle_0}{(1/3)\left\langle({\omega_f}/{\tau_{fp}})p_{\left\langle\mu\right\rangle}p^{\left\langle\mu\right\rangle}\right\rangle_0}\left\langle p^{\left\langle\mu\right\rangle} \phi_f\right\rangle  \Bigg]\frac{\varepsilon_{eq,f}}{n_{eq,f}}
.\end{multline}
We have obtained the above equation by using the following substitutions, 
\begin{eqnarray}
&&\delta f_f=\phi_f f_{eq,f}, \\ 
&&\left\langle  P^{(0)}_1\phi_f\right\rangle=\int \left({\omega_f}/{\tau_{fp}}\right) P^{(0)}_1 \phi_f f_{eq,f}dP, \\ 
&&\left\langle  p^{\left\langle\mu\right\rangle}\phi_f\right\rangle=\int \left({\omega_f}/{\tau_{fp}}\right) p^{\left\langle\mu\right\rangle} \phi_f f_{eq,f}dP
.\end{eqnarray}
The term $u_{\nu}S3^\nu$ can be calculated as follows. 
\begin{multline}\label{zzee}
u_{\nu} S3^\nu=-u_{\nu} \int \Bigg[ \frac{d}{dP} \Bigg[\int \frac{p^\mu u_\mu}{\tau_{fp}}\Big[\delta f_f-
	\frac{\left\langle({\omega_f}/{\tau_{fp}})\delta f_f\right\rangle_0}{\left\langle{\omega_f}/{\tau_{fp}}\right\rangle_0}+
	P^{(0)}_1\frac{\left\langle({\omega_f}/{\tau_{fp}})P^{(0)}_1\delta f_f\right\rangle_0}{\left\langle({\omega_f}/{\tau_{fp}})P^{(0)}_1P^{(0)}_1\right\rangle_0}\\+ p^{\left\langle\mu\right\rangle}\frac{\left\langle({\omega_f}/{\tau_{fp}})p_{\left\langle\mu\right\rangle}\delta f_f\right\rangle_0}{(1/3)\left\langle({\omega_f}/{\tau_{fp}})p_{\left\langle\mu\right\rangle}p^{\left\langle\mu\right\rangle}\right\rangle_0}\Big]dP\Bigg] \int p^{\nu} g_f n^{-1}_{eq,f} (p^\mu u_\mu) f_{eq,f} dP\Bigg] dP \\
 =0.~~~~~~~~~~~~~~~~~~~~~~~~~~~~~~~~~~~~~~~~~~~~~~~~~~~~~~~~~~~~~~~~\Big[{\rm Due ~ to ~ \mathcal O((\delta f_f)^2)}\Big]
\end{multline}
Now, substituting the results of equations \eqref{xxee}, \eqref{yyee} and \eqref{zzee} in eq. \eqref{xyzee}, we get 
\begin{multline}
\Bigg[  \frac{\left\langle({\omega_f}/{\tau_{fp}})P^{(0)}_1\right\rangle_0}{\left\langle({\omega_f}/{\tau_{fp}})P^{(0)}_1P^{(0)}_1\right\rangle_0} \left\langle \omega_f P^{(0)}_1 \phi_f\right\rangle +  \frac{\left\langle({\omega_f}/{\tau_{fp}})p_{\left\langle\mu\right\rangle}\right\rangle_0}{(1/3)\left\langle({\omega_f}/{\tau_{fp}})p_{\left\langle\mu\right\rangle}p^{\left\langle\mu\right\rangle}\right\rangle_0}\left\langle \omega_f p^{\left\langle\mu\right\rangle} \phi_f\right\rangle  \Bigg]\\
-\Bigg[  \frac{\left\langle({\omega_f}/{\tau_{fp}})P^{(0)}_1\right\rangle_0}{\left\langle({\omega_f}/{\tau_{fp}})P^{(0)}_1P^{(0)}_1\right\rangle_0} \left\langle P^{(0)}_1 \phi_f\right\rangle +  \frac{\left\langle({\omega_f}/{\tau_{fp}})p_{\left\langle\mu\right\rangle}\right\rangle_0}{(1/3)\left\langle({\omega_f}/{\tau_{fp}})p_{\left\langle\mu\right\rangle}p^{\left\langle\mu\right\rangle}\right\rangle_0}\left\langle p^{\left\langle\mu\right\rangle} \phi_f\right\rangle  \Bigg]\frac{\varepsilon_{eq,f}}{n_{eq,f}}=0.
\end{multline}
The above equation serves as the matching condition for the novel BGK model and can be expressed as
\begin{equation}
   Q2 ~ n_{eq,f}=Q1 ~ \varepsilon_{eq,f}
,\end{equation}
where, 
\begin{multline}\label{q1}
    Q1=\Bigg[  \frac{\left\langle({\omega_f}/{\tau_{fp}})P^{(0)}_1\right\rangle_0}{\left\langle({\omega_f}/{\tau_{fp}})P^{(0)}_1P^{(0)}_1\right\rangle_0} \left\langle P^{(0)}_1 \phi_f\right\rangle +  \frac{\left\langle({\omega_f}/{\tau_{fp}})p_{\left\langle\mu\right\rangle}\right\rangle_0}{(1/3)\left\langle({\omega_f}/{\tau_{fp}})p_{\left\langle\mu\right\rangle}p^{\left\langle\mu\right\rangle}\right\rangle_0}\left\langle p^{\left\langle\mu\right\rangle} \phi_f\right\rangle  \Bigg]
,\end{multline}
\begin{multline}\label{q2}
    Q2=\Bigg[  \frac{\left\langle({\omega_f}/{\tau_{fp}})P^{(0)}_1\right\rangle_0}{\left\langle({\omega_f}/{\tau_{fp}})P^{(0)}_1P^{(0)}_1\right\rangle_0} \left\langle \omega_f P^{(0)}_1 \phi_f\right\rangle +  \frac{\left\langle({\omega_f}/{\tau_{fp}})p_{\left\langle\mu\right\rangle}\right\rangle_0}{(1/3)\left\langle({\omega_f}/{\tau_{fp}})p_{\left\langle\mu\right\rangle}p^{\left\langle\mu\right\rangle}\right\rangle_0}\left\langle \omega_f p^{\left\langle\mu\right\rangle} \phi_f\right\rangle  \Bigg]
.\end{multline}
In the present scenario, Q1 and Q2 represent the constraints for the conservation of the particle number 
and the energy-momentum, respectively. The consistency of our formulation within the generalised 
hydrodynamic frame described in ref. \cite{Rocha:2021zcw} may be readily observed from equations \eqref{q1} and \eqref{q2}. 

\renewcommand{\theequation}{C.\arabic{equation}}
\section{Derivation of equation (\ref{xx2})}\label{appendix C}
In the BGK model, the relativistic Boltzmann transport equation is written as
\begin{equation}\label{xd3}
q_{f}\textbf{E}\cdot\textbf{p}\frac{\partial f_{eq,f}}{\partial p^0}+q_f p_0\textbf{E}\cdot\frac{\partial f_{eq,f}}{\partial \textbf{p}}=-\frac{p^\mu u_\mu}{\tau_{fp}}\left( f_f(x,p,t)-\frac {n_f(x,t)}{n_{eq, f}}f_{eq,f}(p)\right)
,\end{equation}
where, 
\begin{align}\label{xd4}
\nonumber -\frac{p^\mu u_\mu}{\tau_{fp}}\left( f_f(x,p,t)-\frac {n_f(x,t)}{n_{eq, f}}f_{eq,f}(p)\right) =& -\frac{p^\mu u_\mu}{\tau_{fp}}\left(f_f(x,p,t)-\frac{g_f \int_p (f_{\mathrm eq,f}+\delta f_f)}{n_{\mathrm eq, f}}
f_{eq,f}(p)\right) \\ =& \nonumber -\frac{p^\mu u_\mu}{\tau_{fp}}\left(f_f(x,p,t)-\frac{(g_f \int_p f_{\mathrm eq,f}
+g_f\int_p\delta f_f)}{n_{\mathrm eq,f}}f_{\mathrm eq,f}(p)\right) \\ =& -\frac{p^\mu u_\mu}{\tau_{fp}}\left(\delta f_f-g_f n_{\mathrm eq,f}^{-1}
f_{\mathrm eq,f}\int_p\delta f_f\right)
.\end{align}

Now, following the formulation of ref. \cite{Rocha:2021zcw} in the novel BGK model, we get 
	\begin{multline}\label{x15}
\textbf{C}\bigl[\textit{f}_\textit{f}\hspace{1mm}\bigr]^{NBGK}= -\frac{p^\mu u_\mu}{\tau_{fp}}\Bigg[\delta f_f-
	\frac{\left\langle({\omega_f}/{\tau_{fp}})\delta f_f\right\rangle_0}{\left\langle{\omega_f}/{\tau_{fp}}\right\rangle_0}+
	P^{(0)}_1\frac{\left\langle({\omega_f}/{\tau_{fp}})P^{(0)}_1\delta f_f\right\rangle_0}{\left\langle({\omega_f}/{\tau_{fp}})P^{(0)}_1P^{(0)}_1\right\rangle_0}
  \\
	+ p^{\left\langle\mu\right\rangle}\frac{\left\langle({\omega_f}/{\tau_{fp}})p_{\left\langle\mu\right\rangle}\delta f_f\right\rangle_0}{(1/3)\left\langle({\omega_f}/{\tau_{fp}})p_{\left\langle\mu\right\rangle}p^{\left\langle\mu\right\rangle}\right\rangle_0}\Bigg]
 + g_f n^{-1}_{eq,f} (p^\mu u_\mu) f_{eq,f}\int\frac{p^\mu u_\mu}{\tau_{fp}}\Bigg[ \delta f_f-
	\frac{\left\langle({\omega_f}/{\tau_{fp}})\delta f_f\right\rangle_0}{\left\langle{\omega_f}/{\tau_{fp}}\right\rangle_0}\\
 +
	P^{(0)}_1\frac{\left\langle({\omega_f}/{\tau_{fp}})P^{(0)}_1\delta f_f\right\rangle_0}{\left\langle({\omega_f}/{\tau_{fp}})P^{(0)}_1P^{(0)}_1\right\rangle_0}
	+ p^{\left\langle\mu\right\rangle}\frac{\left\langle({\omega_f}/{\tau_{fp}})p_{\left\langle\mu\right\rangle}\delta f_f\right\rangle_0}{(1/3)\left\langle({\omega_f}/{\tau_{fp}})p_{\left\langle\mu\right\rangle}p^{\left\langle\mu\right\rangle}\right\rangle_0}\Bigg] dP.
\end{multline}
Using  eq. \eqref{x15} and the partial derivatives as given in eq. \eqref{G}, 
eq. \eqref{xd3} takes the following form, 
 \begin{multline}
	\frac{p^\mu u_\mu}{\tau_{fp}}\Bigg[\delta f_f-
	\frac{\left\langle({\omega_f}/{\tau_{fp}})\delta f_f\right\rangle_0}{\left\langle{\omega_f}/{\tau_{fp}}\right\rangle_0}+
	P^{(0)}_1\frac{\left\langle({\omega_f}/{\tau_{fp}})P^{(0)}_1\delta f_f\right\rangle_0}{\left\langle({\omega_f}/{\tau_{fp}})P^{(0)}_1P^{(0)}_1\right\rangle_0} + p^{\left\langle\mu\right\rangle}\frac{\left\langle({\omega_f}/{\tau_{fp}})p_{\left\langle\mu\right\rangle}\delta f_f\right\rangle_0}{(1/3)\left\langle({\omega_f}/{\tau_{fp}})p_{\left\langle\mu\right\rangle}p^{\left\langle\mu\right\rangle}\right\rangle_0}\Bigg]\\
 - g_f n^{-1}_{eq,f} (p^\mu u_\mu) f_{eq,f}\int \frac{p^\mu u_\mu}{\tau_{fp}}\Bigg[ \delta f_f-
	\frac{\left\langle({\omega_f}/{\tau_{fp}})\delta f_f\right\rangle_0}{\left\langle{\omega_f}/{\tau_{fp}}\right\rangle_0} +
	P^{(0)}_1\frac{\left\langle({\omega_f}/{\tau_{fp}})P^{(0)}_1\delta f_f\right\rangle_0}{\left\langle({\omega_f}/{\tau_{fp}})P^{(0)}_1P^{(0)}_1\right\rangle_0}\\
 + p^{\left\langle\mu\right\rangle}\frac{\left\langle({\omega_f}/{\tau_{fp}})p_{\left\langle\mu\right\rangle}\delta f_f\right\rangle_0}{(1/3)\left\langle({\omega_f}/{\tau_{fp}})p_{\left\langle\mu\right\rangle}p^{\left\langle\mu\right\rangle}\right\rangle_0}\Bigg] dP={2q_f\beta(\textbf{E}\cdot\textbf{p})f_{eq,f}(1-f_{eq,f})}.
\end{multline}
For a proper treatment of the left-hand side of the above equation, we are considering an infinitesimal perturbation to the equilibrium distribution function, {\em i.e.} $f_f=f_{eq,f}+\delta f_f$ with $\delta f_f\ll f_{eq,f}$ and after linearizing it, we get 
\begin{equation}
\frac{\omega_f A}{\tau_{fp}} \delta f^{(0)}_f={2q_f\beta (\textbf{E}\cdot\textbf{p}) f_{eq,f}(1-f_{eq,f})}
,\end{equation}
where $ \delta f^{(0)}_f$ is related to the infinitesimal change of the quark distribution function ($\delta f_f$) by the relation given in eq. \eqref{dr1}. After putting the value of  $\delta f^{(0)}_f$ in eq. \eqref{dr1}, we get $\delta f_f$ as given in eq. \eqref{xx2}. 

\renewcommand{\theequation}{D.\arabic{equation}}
\section{Derivation of equation (\ref{mx2})}\label{appendix D}
With the help of equations \eqref{xx}, \eqref{yy} and \eqref{zz}, eq. \eqref{mbx2} can be rewritten as
	\begin{multline}\label{t34}
	p^{\mu}\partial_{\mu}T\frac{\partial f_{eq,f}}{\partial T}+p^{\mu} \partial_\mu(p^{\nu}u_{\nu})\frac{\partial f_{eq,f}}{\partial p^0}+ q_f\Bigr[F^{0i}p_i\frac{\partial f_{eq,f}}{\partial p^0}+F^{i0}p_0\frac{\partial f_{eq,f}}{\partial p^i}\Bigl]\\
	=-\frac{\omega_f}{\tau_{fp}}\biggr[
	P^{(0)}_1\frac{\left\langle({\omega_f}/{\tau_{fp}})P^{(0)}_1\right\rangle_0}{\left\langle({\omega_f}/{\tau_{fp}})P^{(0)}_1P^{(0)}_1\right\rangle_0}\delta f_f\\
	+p^{\left\langle\mu\right\rangle}\frac{\left\langle({\omega_f}/{\tau_{fp}})p_{\left\langle\mu\right\rangle}\right\rangle_0}{(1/3)\left\langle({\omega_f}/{\tau_{fp}})p_{\left\langle\mu\right\rangle}p^{\left\langle\mu\right\rangle}\right\rangle_0}\delta f_f\Biggl]=-\frac{\omega_f A}{\tau_{fp}}\delta f_f	.	
\end{multline}	
Using the following partial derivatives, 
\begin{equation}\label{po9}	
\left.\begin{aligned}
\frac{\partial f_{eq,f}}{\partial T}=-\beta^2 \omega_f f_{eq,f}(1-f_{eq,f}), \\
\frac{\partial f_{eq,f}}{\partial p^0}=-\beta  f_{eq,f}(1-f_{eq,f}), \\
\frac{\partial f_{eq,f}}{\partial \textbf{p}}=-\beta \frac{\textbf{p}}{\omega_f} f_{eq,f}(1-f_{eq,f})
,\end{aligned}
\right\}
\end{equation}
we solve eq. \eqref{t34} to get the expression of $\delta f_f$ as given in eq. \eqref{mx2}. 

\renewcommand{\theequation}{E.\arabic{equation}}
\section{Derivation of equation (\ref{xxx51})}\label{appendix E}
In the BGK model, the relativistic Boltzmann transport equation can be written as
\begin{multline}
	p^{\mu}\partial_{\mu}T\frac{\partial f_{eq,f}}{\partial T}+p^{\mu} \partial_\mu(p^{\nu}u_{\nu})\frac{\partial f_{eq,f}}{\partial p^0}+ q_f\Bigr[F^{0i}p_i\frac{\partial f_{eq,f}}{\partial p^0}+F^{i0}p_0\frac{\partial f_{eq,f}}{\partial p^i}\Bigl]\\=-\frac{p^\mu u_\mu}{\tau_{fp}}\left( f_f(x,p,t)-\frac {n_f(x,t)}{n_{eq, f}}f_{eq,f}(p)\right).
	\end{multline}
With the help of eq. \eqref{xd4}, eq. \eqref{x15} and the partial derivatives as given in eq. \eqref{po9}, we can express the above equation in the novel BGK model as
 \begin{multline}
		\frac{p^\mu u_\mu}{\tau_{fp}}\Bigg[\delta f_f-
	\frac{\left\langle({\omega_f}/{\tau_{fp}})\delta f_f\right\rangle_0}{\left\langle{\omega_f}/{\tau_{fp}}\right\rangle_0}+
	P^{(0)}_1\frac{\left\langle({\omega_f}/{\tau_{fp}})P^{(0)}_1\delta f_f\right\rangle_0}{\left\langle({\omega_f}/{\tau_{fp}})P^{(0)}_1P^{(0)}_1\right\rangle_0} + p^{\left\langle\mu\right\rangle}\frac{\left\langle({\omega_f}/{\tau_{fp}})p_{\left\langle\mu\right\rangle}\delta f_f\right\rangle_0}{(1/3)\left\langle({\omega_f}/{\tau_{fp}})p_{\left\langle\mu\right\rangle}p^{\left\langle\mu\right\rangle}\right\rangle_0}\Bigg]\\
 - g_f n^{-1}_{eq,f} (p^\mu u_\mu) f_{eq,f}\int \frac{p^\mu u_\mu}{\tau_{fp}}\Bigg[ \delta f_f-
	\frac{\left\langle({\omega_f}/{\tau_{fp}})\delta f_f\right\rangle_0}{\left\langle{\omega_f}/{\tau_{fp}}\right\rangle_0} +
	P^{(0)}_1\frac{\left\langle({\omega_f}/{\tau_{fp}})P^{(0)}_1\delta f_f\right\rangle_0}{\left\langle({\omega_f}/{\tau_{fp}})P^{(0)}_1P^{(0)}_1\right\rangle_0}\\
 + p^{\left\langle\mu\right\rangle}\frac{\left\langle({\omega_f}/{\tau_{fp}})p_{\left\langle\mu\right\rangle}\delta f_f\right\rangle_0}{(1/3)\left\langle({\omega_f}/{\tau_{fp}})p_{\left\langle\mu\right\rangle}p^{\left\langle\mu\right\rangle}\right\rangle_0}\Bigg] dP\\=-\frac{f_{eq,f}(1-f_{eq,f})p_0}{T}\Bigr[\frac{p_0}{T}\partial_0 T
	+\Bigr(\frac{p_0-h_f}{p_0}\Bigl)\frac{p^i}{T}\Bigr(\partial_i T-\frac{T}{\varepsilon_f+P_f}\partial_i P_f\Bigl)+T \partial_0\Bigr(\frac{\mu_f}{T}\Bigl)-\frac{p^i p^{\nu}}{p_0}\partial_i u_{\nu}
	-\frac{2 q_f}{p_0}\textbf{E}\cdot\textbf{p}\Bigl].
\end{multline}
After some mathematical calculation, we get 
\begin{multline}
\frac{\omega_f A}{\tau_{fp}} \delta f^{(0)}_f=-\frac{\omega_f f_{eq,f}(1-f_{eq,f})}{T}\Bigr[\frac{p_0}{T}\partial_0 T+\Bigr(\frac{p_0-h_f}{p_0}\Bigl)\frac{p^i}{T}\Bigr(\partial_i T-\frac{T}{\varepsilon_f+P_f}\partial_i P_f\Bigl)\\+T \partial_0\Bigr(\frac{\mu_f}{T}\Bigl)-\frac{p^i p^{\nu}}{p_0}\partial_i u_{\nu}
	-\frac{2 q_f}{p_0}\textbf{E}\cdot\textbf{p}\Bigl],
 \end{multline}
where $\delta f^{(0)}_f$ is related to the infinitesimal change of the distribution function ($\delta f_f$) 
by the relation given in eq. \eqref{dr2}. After putting the value of $\delta f^{(0)}_f$ 
in eq. \eqref{dr2}, the value of $\delta f_f$, {\em i.e.} eq. \eqref{xxx51} is obtained. 

\end{appendix}

\end{document}